\documentclass[10pt, conference, compsocconf]{IEEEtran} 
\IEEEoverridecommandlockouts
\usepackage{mathptmx} 

\usepackage[keeplastbox]{flushend}
\usepackage{fancyhdr}
\usepackage{xcolor}
\usepackage{soul}
\usepackage[hyphens]{url}
\usepackage[bookmarks=true,breaklinks=true,letterpaper=true,colorlinks,linkcolor=black,citecolor=blue,urlcolor=black]{hyperref}
\usepackage{enumitem}
\usepackage{caption}
\usepackage{subcaption}
\usepackage{graphicx}
\usepackage{float}
\usepackage[algoruled,boxed,lined]{algorithm2e}
\usepackage{booktabs, dcolumn}
\usepackage{multirow}
\usepackage[para,online,flushleft]{threeparttable}

\newcolumntype{C}[1]{>{\centering\let\newline\\\arraybackslash\hspace{1pt}}m{#1}}
\newcolumntype{L}[1]{>{\raggedleft\let\newline\\\arraybackslash\hspace{1pt}}m{#1}}

\newcommand{\nsf}[1]{\href{https://www.nsf.gov/awardsearch/showAward?AWD_ID=#1}{#1}}
\usepackage[keeplastbox]{flushend}
\usepackage{cite}
\usepackage{amsmath,amssymb,amsfonts}
\usepackage{algorithmic}
\usepackage{textcomp}
\def\BibTeX{{\rm B\kern-.05em{\sc i\kern-.025em b}\kern-.08em
    T\kern-.1667em\lower.7ex\hbox{E}\kern-.125emX}}
\usepackage{lipsum}
\usepackage{balance}
\fancypagestyle{firstpage}{
  \fancyhf{}
  
  \fancyfoot[C]{\thepage}
}

\begin{document}

\title{Leaky Frontends: Security Vulnerabilities\\ in Processor Frontends} 

\author{
\IEEEauthorblockN{Shuwen Deng, Bowen Huang, and Jakub Szefer}\\
\IEEEauthorblockA{
Yale University\\
\{shuwen.deng, bowen.huang, jakub.szefer\}@yale.edu}\\   
}

\maketitle

\thispagestyle{firstpage}
\pagestyle{plain}

  \begin{abstract}
  This paper evaluates new security threats due to the processor frontend in modern Intel processors. The root causes of the security threats are the multiple paths in the processor frontend that the micro-operations can take: through the Micro-Instruction Translation Engine (MITE), through the Decode Stream Buffer (DSB), also called the Micro-operation Cache, or through the Loop Stream Detector (LSD). Each path has its own unique timing and power signatures, which lead to the side- and covert-channel attacks presented in this work. Especially, the switching between the different paths leads to observable timing or power differences which, as this work demonstrates, could be exploited by attackers. Because of the different paths, the switching, and way the components are shared in the frontend between hardware threads, two separate threads are able to be mutually influenced and timing or power can reveal activity on the other thread. The security threats are not limited to multi-threading, and this work further demonstrates new ways for leaking execution information about SGX enclaves or a new in-domain Spectre variant in single-thread setting. Finally, this work demonstrates a new method for fingerprinting the microcode patches of the processor by analyzing the behavior of different paths in the frontend.  The findings of this work highlight the security threats associated with the processor frontend and the need for deployment of defenses for the modern processor frontend.
  \end{abstract}

  
  \section{Introduction}
\label{sec:introduction}

The processor frontend is responsible for fetching, decoding and delivering micro-ops to
the rest of the processor pipeline.
To achieve efficient decoding and delivery,
multiple paths, and corresponding functional units,
are widely adopted in today's processor designs, such as from Intel~\cite{IntelManual}.
The existence of these multiple paths can lead to security issues, which are explored in this work.

In particular,
we study security of the multiple paths in Intel processor frontend that micro-operations,
also called micro-ops, can take: 
through the Micro-Instruction Translation Engine (MITE), through the Decode Stream Buffer (DSB), 
also called the Micro-op Cache, or through the Loop Stream Detector (LSD).
The LSD was first introduced starting from Intel Core microarchitecture and the
DSB was first introduced starting from Intel SandyBridge microarchitecture
to improve performance and power and to augment the
previously existing MITE.
Due to the existence of the different units, the instruction decoding in a modern processor frontend then has
a unique feature where the same instruction decoding and delivery of
micro-ops can take three different paths: through
MITE, DSB, or LSD. The execution timing and power
depend on the exact path taken in the frontend. In addition, different events can cause switching of the paths based
on the activity in other hyper threads, code loop sizes, or instruction prefixes used,
which further affects timing and power differences.
These different behaviors are basis of the vulnerabilities
that we~demonstrate. 

In this work we demonstrate numerous attacks in both multi-threading (MT) 
and non-multithreading (non-MT) settings.
Unlike majority of existing attacks which happen after the instructions have already been decoded,
our work demonstrates new security problems due to the behavior of the frontend~paths.

Our MT attacks use different threads for the sender and the 
receiver, and leverage evictions or misalignments in DSB or LSD
to create different timing or power variations that can be measured by the receiver. 
For all the covert-channel attacks, the attacks only affect
the frontend and do not, for example, cause interference in the L1 instruction (L1I) caches.
The MT attacks can further be applied to attack SGX enclaves.
We also show MT side-channel attack where the receiver is able to
identify the type of victim application running. The receiver of side-channel attack is
a modified covert-channel receiver that has limited L1I footprint.

We also present attacks that do not require multi-threading.
Our non-MT attacks
mainly use internal-interference among the sender's
own code to cause timing or power variations that the attacker can measure. The non-MT attacks 
can be applied to both SGX or as a new in-domain Spectre attack.

In addition to new attacks on the frontend itself, we demonstrate fingerprinting 
approach that can use frontend behavior
to determine which processor microcode patches have been applied.
Knowing which microcode patches have been applied
can make the attacker stage further attacks by knowing which patch has or has not been applied.

For the different attacks and fingerprinting, timing can be measured by unprivileged attackers.
The power meanwhile can be measured by attackers that can access energy counters, e.g., Intel's Running 
Average Power Limit (RAPL)~\cite{IntelManual}, available in today's processors.
The attacks can thus be done in software and remotely, and part of our evaluation uses public, cloud-based servers
for attack demonstration.

Having presented new microarchitectural vulnerabilities, 
this paper highlights the need to develop protections for the processor frontend.
In particular, the already partitioned DSB and LSB in Intel processors~\cite{IntelManual}
do not provide a full protection as
all our attacks work despite the partitioning.

\subsection{Contributions}

The contributions of this paper are:

\begin{itemize}
 \item Development of the frontend attacks which can covertly send bits between hyper-threads or 
 on the same thread using internal-interference.
 \item Design of both timing-based and power-based variants of the attacks.
 \item Development of attacks leveraging special instruction prefixes to force frontend path switches.
 \item Demonstration of the frontend attacks' ability to leak information from Intel SGX enclaves.
 \item Demonstration of the use of the frontend covert-channels as part of a new Spectre attack variant.
 \item Development of frontend fingerprinting to detect which microcode patch has been applied.
 \item Demonstration of practical frontend-based side-channel used to leak information about victim application type.
\end{itemize}

\subsection{Responsible Disclosure and Open-Source Code}

Our research findings have been shared with Intel.
The code used in this paper will be released under open-source license
at \url{https://caslab.csl.yale.edu/code/leaky-frontends}.

  \section{Background}
\label{sec:background}

In the past, researchers have focused mainly on
attacks leveraging features in the processor backend, while this
work focuses on processor frontend.  Especially, we show timing-based and power-based attacks to bring awareness
that processor frontend needs to be considered when ensuring security of processor architectures.

Previously, security vulnerabilities have been uncovered in all the different levels of 
caches~\cite{bernstein2005cache,gruss2016flush+,yarom2014flush+,liu2015last}, as well as due to 
port contention in the execution engine~\cite{aldaya2019port}, branch 
predictors~\cite{evtyushkin2018branchscope,evtyushkin2016jump},
or memory controllers~\cite{wang2014timing}, for example.
Security community has especially focused on the speculative execution attacks,
following disclosure of Spectre~\cite{Kocher2018spectre} and
Meltdown~\cite{lipp2018meltdown}.
Other recently explored vulnerabilities include attacks that abuse branch prediction, but not for Spectre-like attacks.
This includes BranchScope vulnerability~\cite{evtyushkin2018branchscope} 
or Jump over ASLR type vulnerabilities~\cite{evtyushkin2016jump}.
There are also attacks that leverage prefetchers~\cite{gruss2016prefetch} and value predictors~\cite{deng2021new}.
Most recently, researchers have also demonstrated microarchitectural replay attacks~\cite{skarlatos2019microscope}
and attacks abusing network-on-chip (NoC)~\cite{paccagnella2021lord}.

The vulnerabilities that we present meanwhile focus on the frontend.
To the best of our knowledge, there is one prior work that has explored frontend 
and the micro-op cache (also called the DSB)
for security attacks~\cite{rensee}. 
The work focused on 
studying eviction of DSB and how it can cause timing differences that attackers can exploit.
Compared to the work~\cite{rensee},
we are able to 1) present both eviction-based and misalignment-based attacks
that leverage the DSB, LSD, and MITE, 
2) show new power attacks, 3) evaluate SGX attacks, 4) analyze LSD influence, 5) use frontend
behavior for microcode patch fingerprinting, 6) analyze instruction prefixes causing switching in the frontend paths for new attacks,
and 7) present a new side-channel attack that identifies victim application type.
There is also one concurrent work~\cite{10.1145/3466752.3480079} which focuses on reverse-engineering the operation
of the frontend in Intel and AMD processors.
We believe our work complements existing work by providing new attacks
and security insights, including, to the best of our knowledge, fastest
frontend attack reaching~$1.4$Mbps.

  \section{Threat Model and Assumptions}
\label{sec:threat_model}

We assume there is one sender (victim) that holds security-critical information and 
one receiver (attacker) that tries to extract the secret information by measuring timing or power changes.
For covert-channels, the sender and receiver cooperate and modulate usage of the DSB, LSB, and MITE 
to achieve the covert transmission.
For side-channels, the attacker performs operations to interfere with the
victim or monitor power or timing,
while the victim is unaware of the attacker and operates on sensitive data.
Our attack on SGX
assumes that the attacker can trigger execution of the enclave and measure its timing or power.
Our Spectre attack
assumes an in-domain attack scenario: the attacker is within same thread, e.g.,
as a sandboxed code where the disclosure gadget is executed.
Our fingerprinting attack
assumes attacker has prior access to the same CPU as the target one, so they can
measure frontend performance under different microcode patches.
All of the timing-based attacks can be performed fully from the user-level privilege using the {\tt rdtscp} instruction
for measuring timing.
The power channels require access to Intel's RAPL~\cite{IntelManual} to get energy information. Even if
the RAPL access is disabled for user-level code,
privileged code can still use the power channels against SGX enclaves, for example.
  
  \section{Analysis of the Operation of the Frontend}
\label{sec:create_channel}

\begin{figure}[!t]
 \centering
 \includegraphics[width=8.1cm]{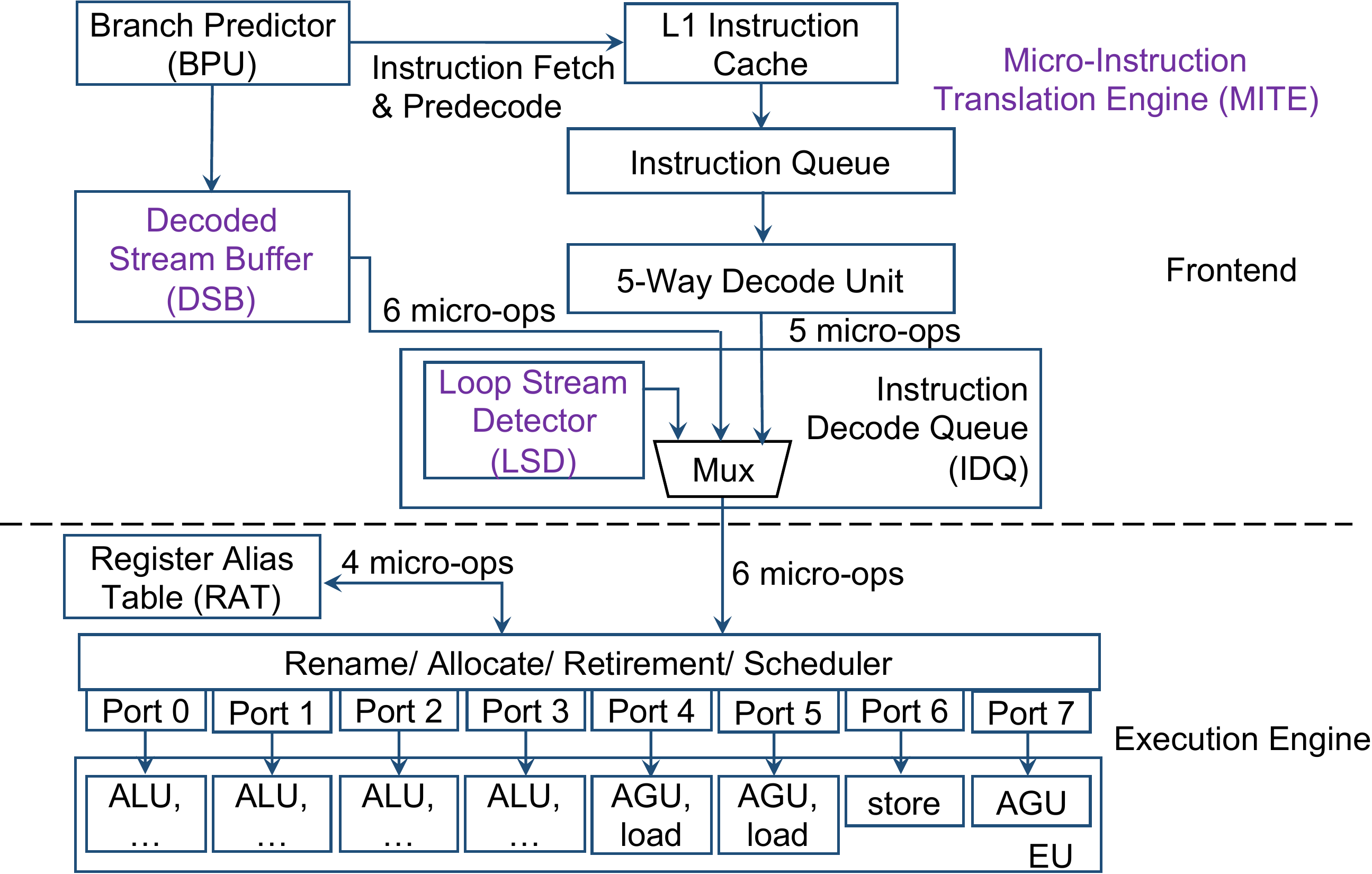}
 \caption{\small Microarchitecture details of the frontend and the execution engine, based on~\cite{IntelManual}.}
 \label{fig:microarchitecture}
\end{figure}

Within the processor frontend, instruction decoding and delivery to the backend
has multiple paths: through the Micro-Instruction Translation 
Engine (MITE), the Decoded Stream Buffer (DSB), also called the micro-op cache,
and the Loop Stream Detector (LSD), as is seen from Figure~\ref{fig:microarchitecture}.

Given that MITE path has low throughput and high power consumption,
the DSB has been added and 
the micro-ops decoded by MITE are inserted into the DSB~\cite{IntelManual} in modern Intel processors.
If the micro-ops are available in the DSB, the micro-op stream
is sent directly from DSB to the Instruction Decode Queue (IDQ),
bypassing the MITE, therefore saving power and improving throughput. 
The instruction delivery path from DSB is also shorter than MITE (shorter by $2$ -- $4$ cycles), 
so the pipeline latency is reduced as well~\cite{IntelManual}.

Further, there is also the LSD located within the IDQ.
If the micro-op stream belongs to a qualified loop (discussed in Section~\ref{subsec:instruction_mix}),
all the micro-ops of the loop code can be issued directly from LSD to the backend,
bypassing DSB as well. 
The purpose of the LSD is to help save power, but it also can help performance
by providing higher instruction delivery throughput.
When branch mis-prediction occurs, e.g., at the end of the loop,
or the number of micro-ops within the loop exceeds the limit that the LSD can handle, 
LSD is not used and micro-ops are delivered from the DSB.
Furthermore, if the micro-ops exceed the DSB limit or belong to a newly accessed micro-ops,
they are processed by the MITE.
We also note that the DSB is inclusive of LSD, and MITE is inclusive of DSB as well~\cite{IntelManual}, e.g.,
eviction of micro-ops from DSB will cause their eviction from LSD.
Although DSB and LSD are partitioned in Intel processor when two hyper-threads are actively running,
our analysis indicates that DSB in Intel processors is fully assigned to one thread if the other is idle or not executing.
When the second thread becomes active, DSB becomes partitioned, which forces
DSB evictions of micro-ops of the first thread to occur.
Further, MITE is a shared resource, and activity of two threads mutually affects the  micro-op decoding.

\subsection{LSD Behavior}

The LSD can continuously stream the same sequence of up to $64$ micro-ops,
directly from the IDQ to the backend~\cite{IntelManual}.
While the LSD is active, the rest of the frontend is effectively disabled.
In order to generate detectable timing and power difference between LSD vs. DSB and DSB vs. MITE, 
one can control micro-op number within
a loop to either make it fit in the LSD where instruction delivery starts with LSD only,
or exceed the LSD limit so the processor falls back to use DSB or MITE, 
creating detectable timing and power~changes.

\subsection{DSB Behavior}

The DSB is constructed as a cache-like structure with $32$ sets and $8$ ways per set~\cite{IntelManual}. 
Each line can store up to $6$ micro-ops or $32$ bytes (so DSB can hold at most $1536$ micro-ops in total). 
Based on our reverse engineering as well as Intel manuals~\cite{IntelManual},
we find that 
when there is only one thread running on the hardware core, 
instructions' virtual address bits {\tt addr[4:0]} are used as the byte offset within the $32$-byte window, 
and {\tt addr[9:5]} are the set index bits into the $32$ DSB sets.
However, when two threads are running in parallel on the hardware core,
the DSB is set partitioned, and half the sets are assigned to each thread based on our experimental results.
This means that 
although the DSB is partitioned by sets when two threads are running,
if there is only one thread being active, the thread is assigned to all the DSB sets.
Whether the DSB is currently partitioned or not can be detected by an application
by checking the increased MITE usage (when DSB is partitioned, more instructions will conflict with each other
causing DSB evictions and increased MITE usage).
The behavior was tested
on Intel Xeon E-2174G with 
LSD disabled to show the conflicts are not influenced by LSD.
We also tested on Intel Gold 6226 with LSD enabled, and observe similar results.
Further, we tested Intel Gold 6226 with LSD enabled, but each test thread was set to access
larger blocks of instructions which do not fit in
LSD (forcing processor to use DSB even if LSD is enabled), and similar results are observed.

\subsection{MITE Behavior}

Regarding the MITE structure, the instruction cache, instruction queue, 
and the decode unit are shared among the two threads.
Typically the instruction cache is $32$ KB and $8$-way associative and instruction queue contains $50$ entries. 
The DSB, LSD and MITE behaviors were tested on Intel processors shown in Table~\ref{tbl:CPU_info}.
These same processors are also used for evaluation of our side- and covert-channel~attacks.

\begin{table}[t]
 \caption{\small Specifications of the tested Intel CPU models.}
 \label{tbl:CPU_info}
 \centering
 \fontsize{8pt}{10pt}\selectfont
 \begin{threeparttable}
 \begin{tabular}{|p{0.85in}|p{0.36in}|p{0.4in}|p{0.4in}|p{0.4in}|}
 \hline
 \textbf{{Model}} & \textbf{{Gold 6226}} & \textbf{{Xeon E-2174G}} & \textbf{{Xeon E-2286G}} & \textbf{{Xeon E-2288G}} \\
 \hline
 Microarchitecture &Cascade Lake &\multicolumn{3}{c|}{Coffee Lake}\\
 \hline
 Core Number & 12 & 4 & 6 & 8\\
 \hline
 Thread Number & 24 & 8 & 12 & 8$^a$\\
 \hline
 L1D Configuration &\multicolumn{4}{c|}{32KB, 8-way, 64 byte line size, 64 sets}\\
 \hline
 DSB Configuration &\multicolumn{4}{c|}{8-way, 32 byte window, 32 sets}\\
 \hline
 LSD Entries & 64 & --- $^b$ & --- $^b$ & 64 \\
 \hline
 Frequency & 2.7GHz & 3.8GHz & 4.0GHz & 3.7GHz\\
 \hline
 OS & \multicolumn{4}{c|}{18.04 Ubuntu} \\
 \hline
 SGX Support & No & \multicolumn{3}{c|}{Yes} \\
 \hline
 \end{tabular}
 \begin{tablenotes}
 \small{$^a$ We use Xeon E-2288G on Microsoft Azure cloud, this processor
 model is specific for Microsoft Azure and has hyper-threading disabled, although
 hyper-threading is supported by other E-2288G processors.}
 \small{$^b$ LSD is disabled in these machines.}
 \end{tablenotes}
 \end{threeparttable}
\end{table}

\subsection{Ensuring Observability of Frontend Timing}
\label{subsec:instruction_mix}

To achieve high backend throughput so that the frontend is the bottleneck,
we do not want to touch data-related operations such as load and store because memory system 
may cause unpredictable timing differences, which are not due to frontend path changes.
Load and store operations would also 
likely leave traces in the caches which may make any attacks more detectable.
Based on our analysis,
instruction mix sequence
which maximizes the timing signature of the frontend for our 
attacks should satisfy the following three~requirements:

\begin{itemize}
 \item Total bytes of 
 one access block should not exceed a 32 byte window 
 (e.g., $4$ \textit{mov} and $1$ \textit{jmp} use in total $25$~bytes). 
 \item Total micro-op number should not exceed $6$ micro-op limit that DSB can process by one DSB line
 (e.g., $4$ \textit{mov} and $1$ \textit{jmp} are decoded to total $5$ micro-ops).
 \item Avoid port contention. The $4$ \textit{mov} instructions exploit the ports as much as possible, plus $1$
 \textit{jmp} instruction to end the cache line block, while 
 avoiding load, store, or more complex instructions involved, which will cause influence 
 or noise from other microarchitectural units.
\end{itemize}

\noindent As the result, $4$ {\em mov} plus $1$ {\em jmp} sequence is the {\em instruction mix block} 
which fits the requirement.
Other instruction mix blocks are possible,
although finding sufficient type and number of instruction mix blocks in
real code may be a limitation of the proposed attacks.

\subsection{Exploiting Frontend Path Timing Differences}

\begin{figure}[!t]
 \centering
 \includegraphics[width=5cm]{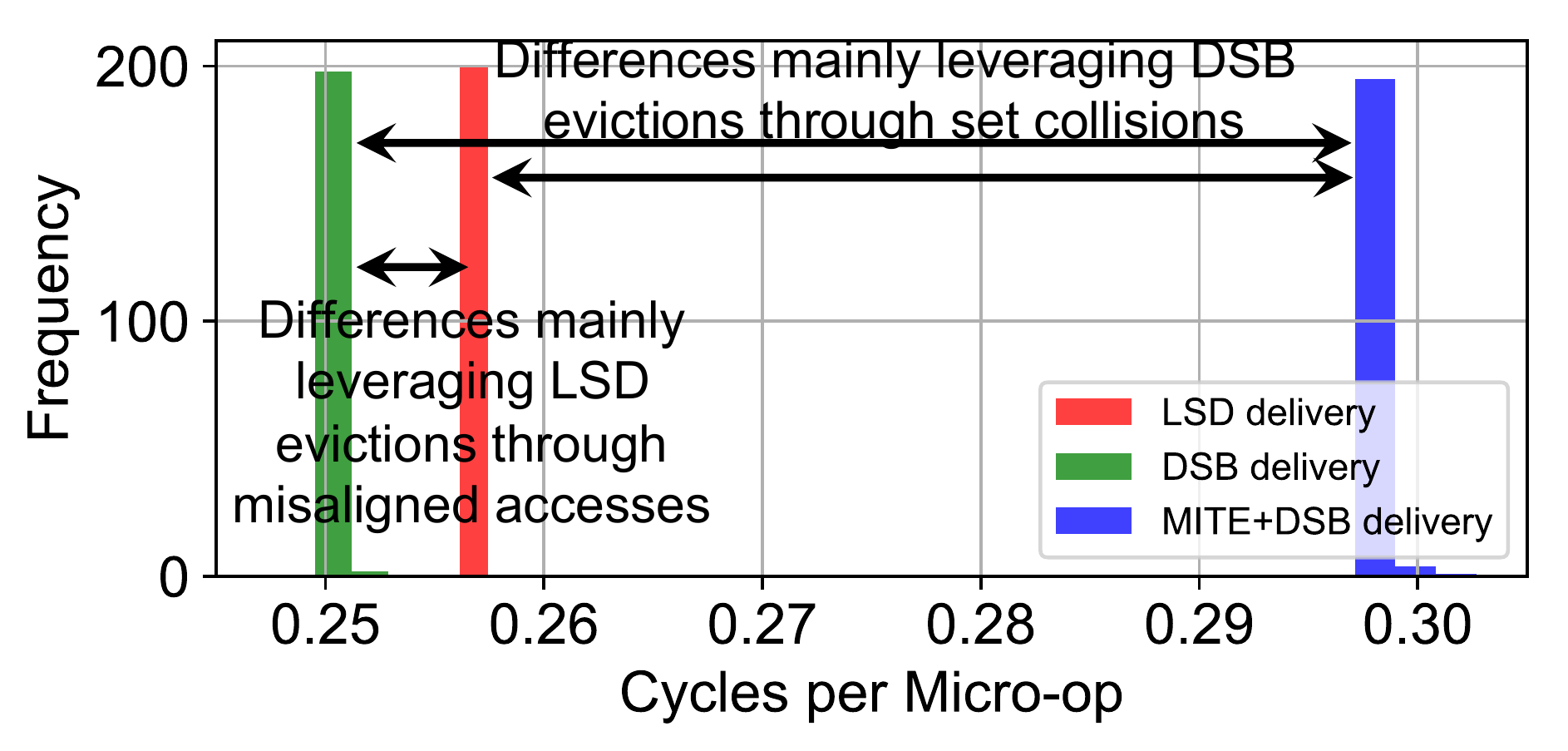}
 \caption{\small Example time histogram of Intel Xeon Gold 6226 processor
 of using LSD, DSB, or MITE+DSB paths. Timing difference between LSD/DSB and MITE+DSB are 
 used for collision-based attacks (see Section~\ref{subsec:covert_channel}) 
 and differences between LSD and DSB paths are used for 
 misalignment-based attacks (see Section~\ref{subsec:misalignment_impl}).}
 \label{fig:leverage}
\end{figure}

As can be seen from histogram of Intel Xeon Gold 6226 processor shown in Figure~\ref{fig:leverage}, 
the timing difference of processing instruction mix blocks
using DSB vs. MITE+DSB or LSD vs. DSB are clearly visible.
In our attacks discussed later, we will use DSB vs. MITE+DSB timing differences to perform attacks 
related to DSB evictions through set collisions.
On the other hand, the timing difference of processing using LSD vs. DSB will be used to 
perform attacks related to LSD evictions through misaligned accesses. 
Both of these also have power differences that separately can be used for power-based~attacks.

\subsection{Generating DSB Evictions Through Set Collisions}
\label{subsec:collision}

\begin{figure}[!t]
 \centering
 \includegraphics[width=8.7cm]{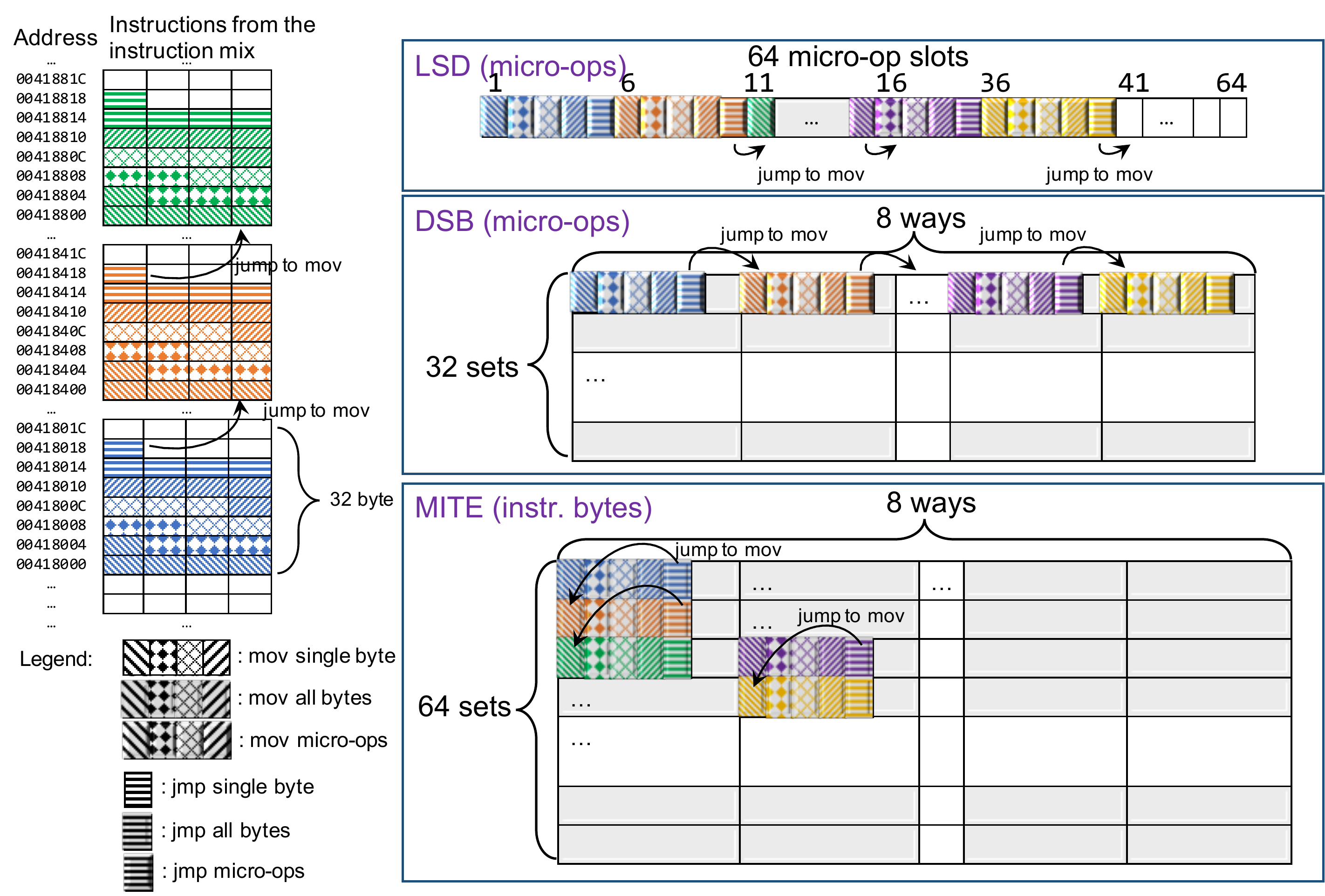}
 \caption{\small Example of mapping instruction mix blocks 
 (Section~\ref{subsec:instruction_mix}) to MITE, DSB, and LSD. 
 Each instruction mix block is $5$ micro-ops ($4$ \textit{mov} plus $1$ \textit{jmp}). 
 If the number of chained $5$ micro-op blocks is $8$ then all will fit in LSD 
 (since $8\times 5=40<64$ micro-op limit of LSD) 
 and they can all map to the same DSB set (since DSB is $8$-way associative).}
 \label{fig:mapping}
\end{figure}

To force frontend path changes, we set up a series of instruction mix blocks 
and align the start of the instruction address of each block 
to map to the same DSB set, as shown in Figure~\ref{fig:mapping}.
We make the {\textit{jmp}} instructions at the end of each instruction mix block jump to the first 
instruction
of next instruction mix block.
In this case, executing the first \textit{mov} instruction
of the first instruction mix block will trigger a series of instruction mix block execution.
If the chain of instruction mix blocks is less than $12$, all the blocks should fit in LSD. 
However, at the same time, each DSB set has $8$ ways, so $8$ blocks can map to same set. 
Consequently, if the chain of blocks is set to $8$ (rather than $12$), they can both fit in LSD and same DSB set. 
But, as soon as the chain is extended to~$9$ (or more) blocks that map to same set, eviction occurs in DSB, 
and in turn force LSD eviction due to inclusive nature of MITE, DSB, and LSD.

Inclusive feature of MITE, DSB, and LSD makes
eviction of lines from DSB to cause flush of 
the LSD unit. 
Furthermore, eviction from DSB redirects micro-ops to be processed by MITE.
Combing these, eviction from DSB will cause transition of micro-op delivery from LSD to both DSB and MITE.

Note that changing the chain of instruction mix blocks from $8$ to $9$ will not
cause eviction or misses of L1 instruction cache.
L1 instruction caches for the machines we tested are $8$-way associative and contain $64$ sets of $64$ bytes.
Consequently, the size of the L1 instruction is $4$ times of DSB and instruction 
mix blocks mapping to the same DSB set will be mapped to different
L1 instruction cache sets, as is shown in Figure~\ref{fig:mapping}.
In other words, changing chain length from $8$ to $9$ causes DSB and LSD eviction, 
but causes no misses in the L1 instruction cache.

\subsection{Generating LSD Evictions Through Misaligned Accesses}
\label{subsec:misalignment}

We further found that misaligned instructions will generate collisions in the LSD,
even when the number of total accessed instruction mix blocks does not exceed the DSB way number.
This can be achieved by setting up the initial addresses of instruction mix blocks 
to be misaligned, e.g., 
by aligning them on $16$ byte boundaries that are not multiple of $32$~bytes.

The alignment or misalignment of the blocks will cause different frontend path changes when processing micro-ops.
When all the instruction mix blocks are misaligned,
executing $4$ chained instruction mix blocks that map to the same DSB set 
will trigger collisions in LSD
which causes the micro-op delivery change from LSD to DSB.
At the same time, as we discussed in Section~\ref{subsec:collision}, 
executing $4$ chained aligned instruction mix blocks that map to the same DSB set will use LSD unit 
since the size of the $4$ blocks (of $5$ micro-ops each) is less than $64$ micro-op limit of the LSD.

When considering accessing pattern,
if accessing 
a chain of~$7$ instruction mix blocks which are all aligned, the $8^{th}$ access will determine the path used.
If the $8^{th}$ access is aligned, all of the micro-ops will still be processed by the LSD.
While if the $8^{th}$ instruction mix block is misaligned, LSD will be flushed and 
micro-ops will be redirected to use DSB in the frontend.
We found that \{aligned + misaligned\} instruction mix block access pairs that will cause 
micro-ops to be changed from the LSD to the DSB paths are:
\{5 aligned + 2 misaligned\}, \{6 aligned + 2 misaligned\},
\{3 aligned + 3 misaligned\}, \{4 aligned + 3 misaligned\}, and \{5 aligned + 3 misaligned\}.
Similar to DSB evictions, misalignment will not cause L1 instruction cache misses.

\begin{figure}
 \centering
 \includegraphics[width=\linewidth]{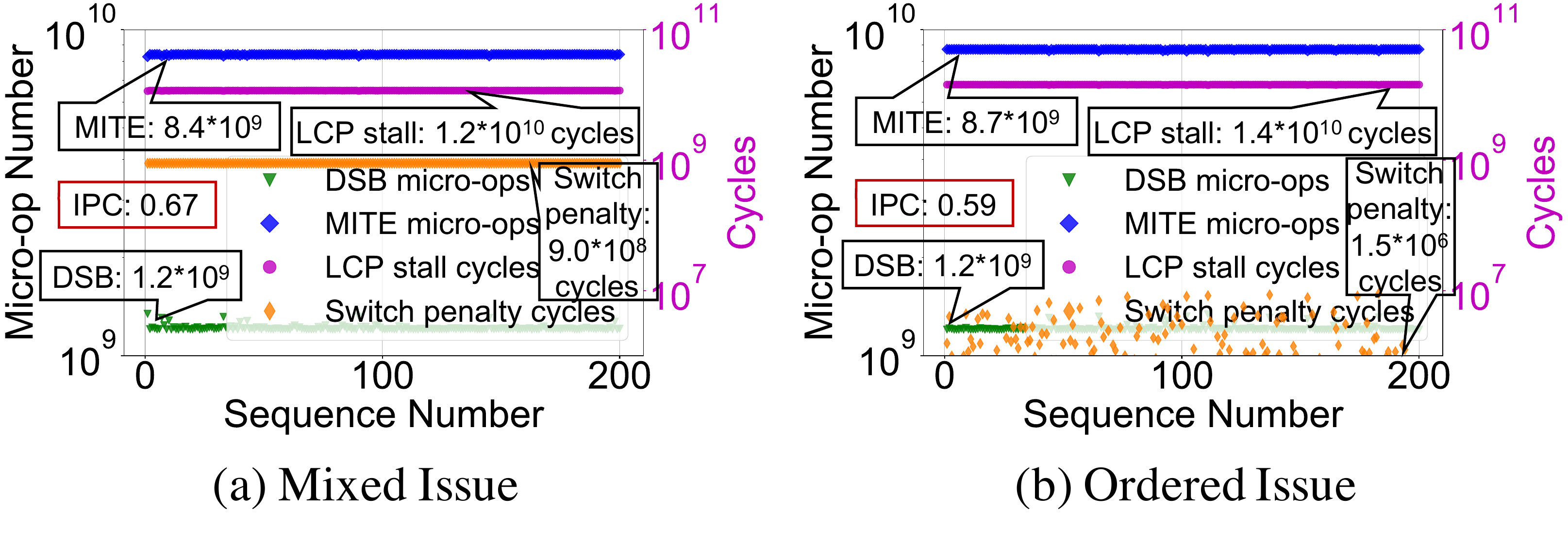}
 \caption{\small Intel Xeon Gold 6226 CPU 
 performance counter readings for the different experiments with ordered-issued 
 or mixed-issued types of \textit{add}
 instructions.
 The numbers in the call-out boxes are the average micro-ops numbers for all the $200$ rounds of experiments.}
 \label{fig:srv3_counter}
\end{figure}

\subsection{Generating Different DSB Switch Penalties}
\label{subsec:dsbswitchpenalty}

In x86, Length Changing Prefixes (LCPs) are designed and incorporated into the x86 ISA to identify 
the instructions with non-default length, which may be used, e.g.,
with unicode processing and image processing~\cite{IntelManual}.
For example, an instruction starting with \textit{0x66h} prefix means there would be an operand size override. 
Such prefix can force CPU to use slower decoding MITE path
and incur up to $3$ cycles more penalty in addition to extra DSB-to-MITE switch~penalty. 

To demonstrate that generating different switch penalties is feasible, we set up two instruction mix blocks. 
The first instruction mix block is filled by $16$ sets of two \textit{add} instructions,
with one normal \textit{add} instruction followed by 
one \textit{add} instruction with length changing prefixes (mixed issue),
and repeating this alternating pattern to the end. 
The second one is filled with $16$ normal \textit{add} instructions followed by 
$16$ \textit{add} instructions with length changing prefixes (ordered issue),
In both cases there are {\tt 32} instructions within the loop and we iterate the loop for {\tt 800} million times.
Figure~\ref{fig:srv3_counter} shows the results of the measurement. 
The two instruction blocks generate similar number of micro-ops from MITE and DSB, 
but with detectable difference in the final performance (measured in instructions per cycle, or IPC), 
which is caused by different numbers of LCP stall cycles and DSB-to-MITE switch penalty cycles. 
This shows that the same type of frequently-used instructions can come with
different front-end path switching penalties.

We also found other possibly useful, for an attacker, LCP behaviors including: 
a) use of LCP will force the front-end to switch from issuing instructions from DSB to issuing instructions from MITE, 
b) LCP instructions are only decoded sequentially and would incur measurable performance difference. 
Therefore, it is feasible to establish a covert channel based on instructions with LCPs.

  \section{Processor Frontend Vulnerabilities}
\label{sec:impl}

In this section, we focus on implementation of the timing-based covert channels and attacks.
Evaluation of the timing-based channels is in Section~\ref{section_evaluation}.
Power-based channels and attacks are discussed in Section~\ref{sec:power_attack}.
Meanwhile, application of the attacks to SGX enclaves is presented in Section~\ref{sec:sgx},
and for use with Spectre attacks in Section~\ref{sec:spectre_attack}.
Detection of the microcode patches, which can use both timing or power, is presented in Section~\ref{sec:patch}.
Finally, a new side-channel attack used to fingerprint applications is in Section~\ref{sec_fingerprinting}.

Our timing-based covert-channel attacks can be differentiated based on
the techniques used to covertly send different bits by switching between different frontend paths:
using eviction (following ideas in Section~\ref{subsec:collision}),
using misalignment (following ideas in Section~\ref{subsec:misalignment}),
or using LCP stalls and DSB-to-MITE switch penalties (following ideas in Section~\ref{subsec:dsbswitchpenalty}).

For our attacks, there are generally three steps that the attacks follow:

\begin{itemize}
 \item \textbf{Init Step}: A series of instruction accesses are performed in this step 
 to set the micro-ops into certain frontend paths,
 for some attacks no initial step is needed, only start timing (or power) measurements.
 \item \textbf{Encode Step}: The sender accesses certain instructions to change frontend 
 paths of micro-ops previously set in the 
 initialization step according to the secret bit to be~sent.
 \item \textbf{Decode Step}: The receiver accesses certain instructions and, depending on attack type,
 timing or power is measured to observe what changes occurred in the frontend, or for all three steps.
\end{itemize}

\noindent
In addition, some of the attacks may require timing or power measurements in not just the last step,
but the attacks still follow the three-step pattern.

In the attack descriptions we use the following variables to describe parameters of the system:
$N$ is the number of ways in the DSB.
$m$ is a 1-bit message to be transferred on the channel.
$d$ is the different number of instruction mix blocks used for an attack step, $d < N+1$.
$M$, only used for misalignment-based attacks, is the parameters of the receiver, $M < N+1$.
$p$ is the number of iterations the receiver runs for initialization step and also for decoding step.
$q$ is the number of iterations the sender runs for encoding step.
We note that use of multiple iteration increases time, but helps to reliably observe timing result with a low error rate.
$r$, only used for attacks leveraging LCP, is the number of LCP instructions.

\subsection{Eviction-Based Timing Attack with Multi-Threading}
\label{subsec:covert_channel}

For the eviction-based attack, in a multi-thread (MT) setting,
we deploy a sender thread and a receiver thread on the same physical processor core,
but different hardware threads, 
which causes them to share the frontend.
When the instruction stream from the sender executes, the DSB will be partitioned and
some of the receiver's instructions will be evicted from DSB, further triggering eviction from LSD so 
that the delivery of instructions falls back from the LSD to DSB+MITE,
therefore generating detectable timing signature that the receiver can measure.
When the instruction stream from the sender is not executing, the receiver thread will use whole DSB
and the evictions will not happen. 
This process leaves no interference in traditional instruction and data caches.

\begin{figure}[!t]
 \centering
 \includegraphics[width=8.6cm]{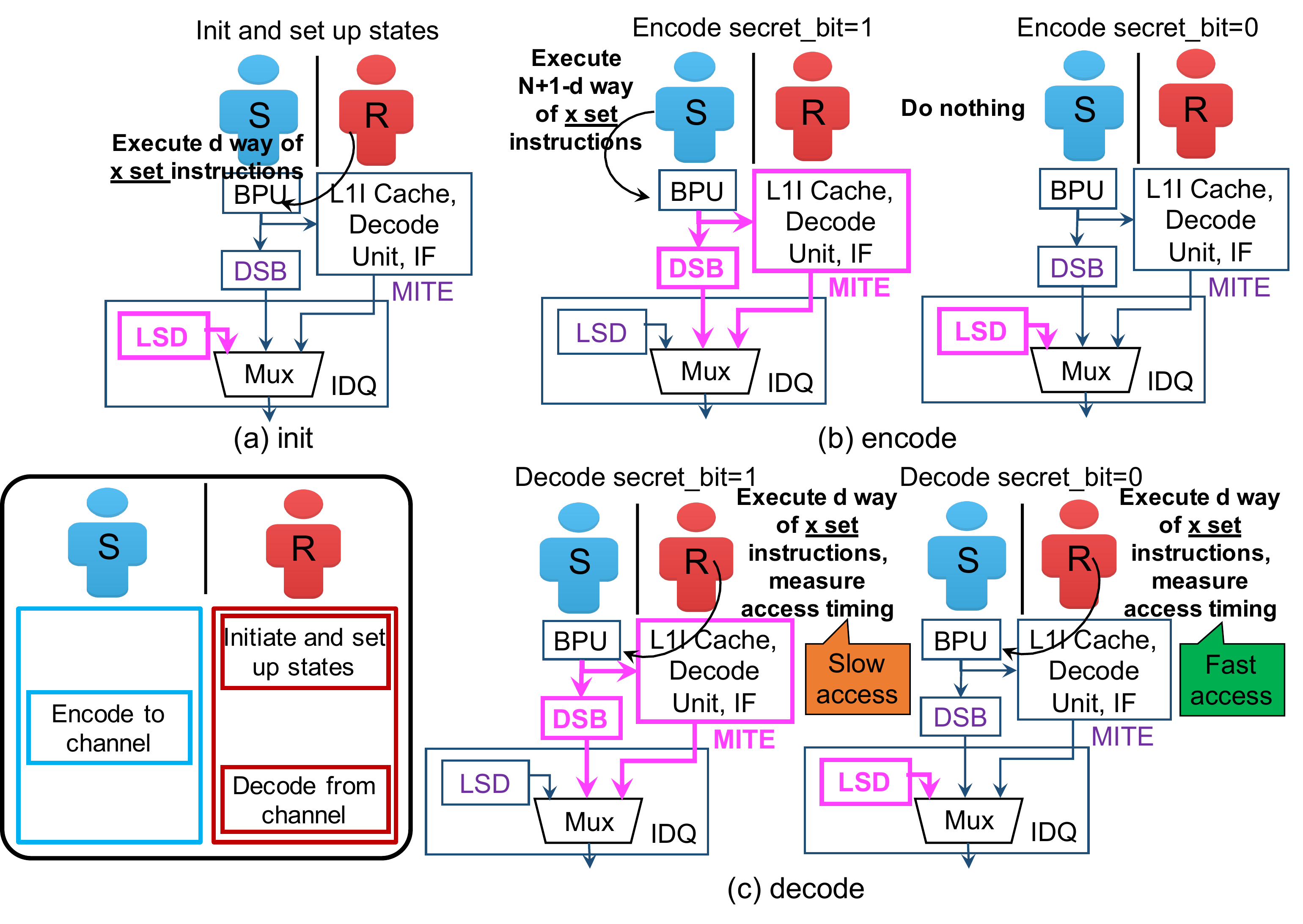}
 \caption{\small Overview of the MT Eviction-Based Attack.}
 \label{fig:impl_2t_evict}
\end{figure}

In the MT Eviction-Based Attack,
the sender and the receiver use in total
$N+1$ instruction mix blocks, denoted as lines $0$ - $N$.
For the MT eviction-based attack shown in Figure~\ref{fig:impl_2t_evict},
in the Init Step,
$d$ ($d \leq N$)
instruction mix blocks that map to a DSB set $x$ are accessed for $p$ times by the receiver. 
In the Encode Step,
the sender will execute different instruction series according to the secret bit $m$.
When sending $m=1$, the sender will execute $N+1-d$
instruction mix blocks $q$ times, these blocks map to DSB set $x$. 
In this case, the total number of ways accessed is larger than $N$, which causes eviction of DSB within the receiver
and directs the micro-op delivery from LSD to DSB and MITE.
When sending $m=0$, the sender does nothing.
In the Decode Step,
the receiver will access the same $d$ instruction mix blocks accessed in the Init Step
and time the Decode Step's
access for $p$ iterations. 
If eviction occurs, receiver's micro-ops in the Decode Step
will be delivered from DSB and MITE, where longer timing
will be measured, indicating message $m=1$ was sent from the sender.
On the other hand, if no evictions happen,
receiver's micro-ops in the Decode Step
will still be delivered from LSD, where much shorter timing is observed compared to the MITE+DSB path, 
indicating message $m=0$ was sent from the sender.

For example, take $d=6$ and $N=8$,
the instruction access sequences when sending $m=1$ and $m=0$ are as~follows:
\begin{itemize}
 \item \textbf{Init}: access blocks $1 - 6$ mapping to set $x$
 \item \textbf{Encode}: access blocks $7 - 9$ mapping to set $x$ (if $m=1$); no access (if $m=0$)
 \item \textbf{Decode}: access blocks $1 - 6$ mapping to
 set $x$ (if $m=1$, DSB and MITE are used; if $m=0$, LSD access is used)
\end{itemize}

\subsection{Misalignment-Based Timing Attack with Multi-Threading}
\label{subsec:misalignment_impl}

To achieve misaligned instruction access,
sender and receiver first find virtual addresses of instructions that map to the same target set as what 
eviction-based attacks do, and then 
offset the initial address of every instruction mix block by $16$ bytes (half of the DSB line size), 
to misalign the address.

For this type of attack, the total number of instruction mix blocks of the sender and the receiver
are equal to or less than the $N$ ways of the DSB, which has an advantage 
as it reduces the number of accesses and increases the transmission rate compared with eviction-based attacks.

\begin{figure}[!t]
 \centering
 \includegraphics[width=8.6cm]{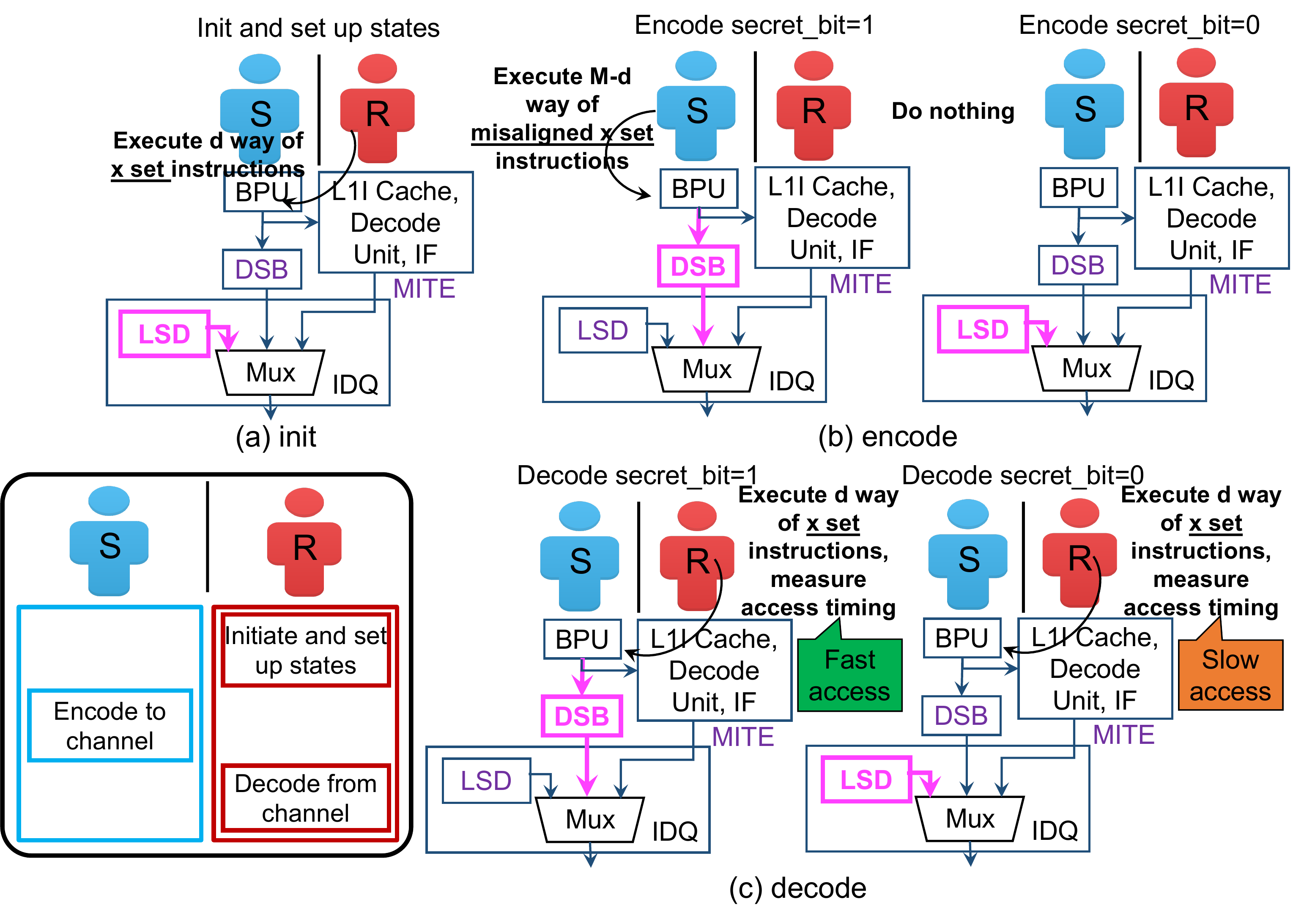}
 \caption{\small 
 Overview of the MT Misalignment-Based Attack.}
 \label{fig:impl_2t_misalign}
\end{figure}

The MT Misalignment-Based Attack is shown in Figure~\ref{fig:impl_2t_misalign}.
Here,
the sender and the receiver use in total
$M$ ($M \leq N$) instruction mix blocks.
In the Init Step and the Decode Step,
the receiver will access in total 
$d$ (where $d<N$) sets of 
instructions mix blocks that map to one DSB set, this is repeated for $p$ times.
In this case, the receiver's instructions accessed in the Init Step
will be processed by the LSD.
For the sender, in the Encode Step,
when sending $m=1$, the sender will execute ($M-d$) (where $M < N+1$) sets of misaligned
instructions that map to the same DSB set as the receiver for $q$ iterations. 
In this case, misalignment of the DSB causes the micro-op delivery to 
be redirected to DSB from LSD,
which leads to faster access of receiver's instruction in the Decode Step.
When sending $m=0$, the sender does nothing.
In this case,
all the micro-ops will still be delivered by the LSD
and the receiver's instruction access in the Decode Step
will observe slower access time.
We note that LSD is indeed slower in delivery which is demonstrated by the evaluation shown in 
Figure~\ref{fig:leverage}.

For example, take $d=5, N=8, M=8$, the access sequences when sending $m=1$ and $m=0$ are as~follows:

\begin{itemize}
 \item \textbf{Init}: access instruction mix blocks $1 - 5$ mapping to set $x$
 \item \textbf{Encode}: access \textit{misaligned} instruction mix 
 blocks $6 - 8$ mapping to set $x$ (if $m=1$); no access (if $m=0$)
 \item \textbf{Decode}: access blocks $1 - 5$ mapping to set $x$ (if $m=1$, DSB 
 access is used; if $m=0$, LSD access is used)
\end{itemize}

\subsection{Non-MT Eviction-Based Attack without Multi-Threading}
\label{subsec:singlethread}

\begin{figure}[!t]
 \centering
 \includegraphics[width=8.6cm]{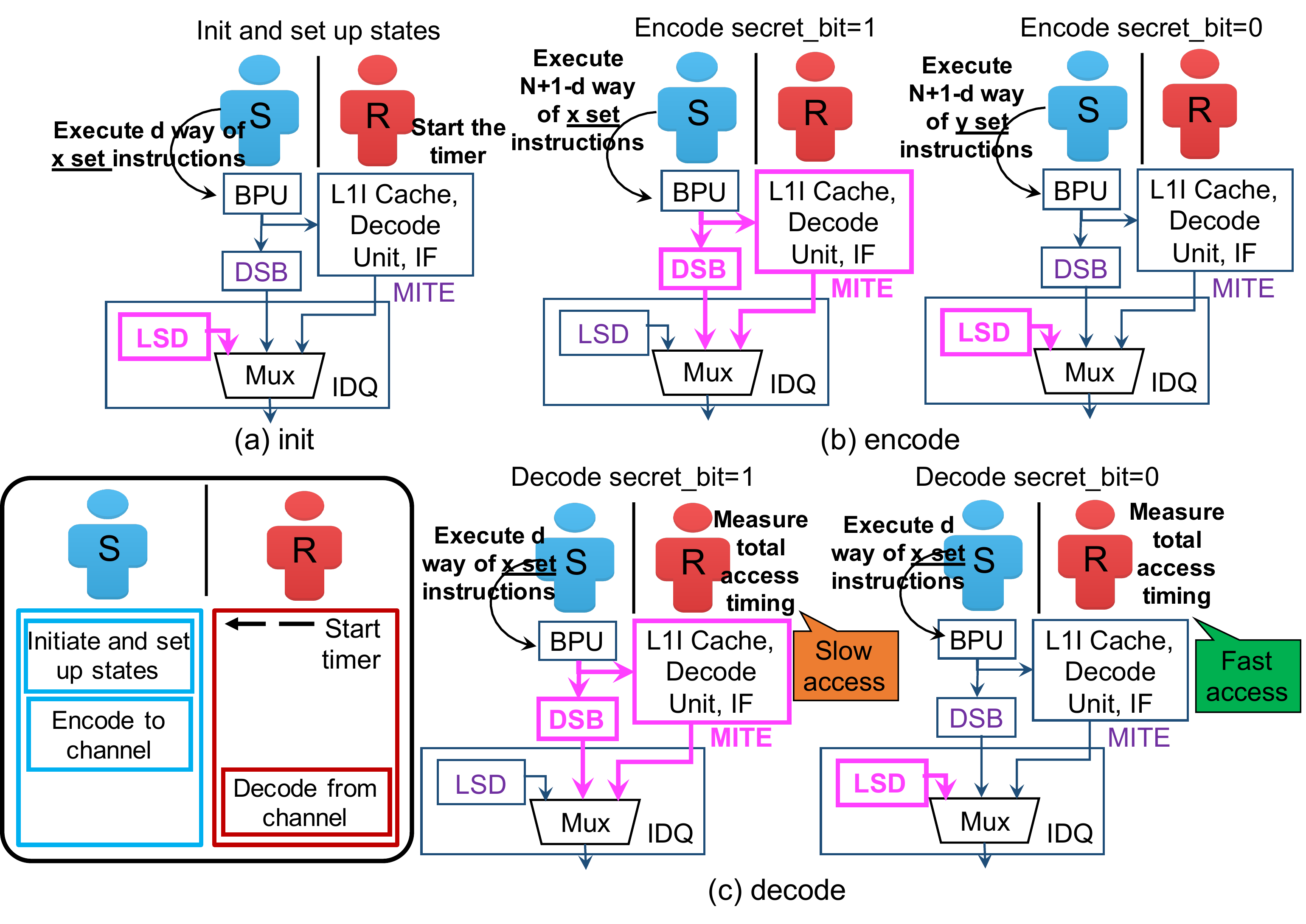}
 \caption{\small Overview of Non-MT Stealthy Eviction-Based Attack.}
 \label{fig:impl_II_evict}
\end{figure}

Our attack using internal-interference of the sender is shown in Figure~\ref{fig:impl_II_evict}.
The number of iterations ($q$) of
sender's encoding step and number of iterations ($p$) of receiver's initialization and decoding steps will be the same
in this attack (i.e. $p=q$)
in order to reliably observe one timing result with a low error rate.
For one iteration, in the Init Step,
the receiver starts the timer in order to measure total time of the sender.
The sender then executes $d$ ($d \leq N$)
instructions mix blocks that map to DSB set $x$. 
The instructions
will be processed by the LSD.
In the Encode Step,
When sending $m=1$, the sender will execute $N+1-d$ 
instruction mix blocks that map to the same DSB set as the receiver. 
When sending $m=0$,
the sender will execute the same number of instruction mix blocks
but ones that map to a different DSB set $y$.
(stealthier for security) 
or do nothing (faster for bandwidth).
In the Decode Step,
the sender will
access the same number $d$ of instruction mix blocks
accessed in the Init Step.
Then the receiver will end the timer and calculate the total timing of the sender's accesses to derive the information sent. 
If the Encode Step's
access 
causes evictions,
sender's micro-ops in the Decode Step
will be delivered from DSB and MITE, where longer timing
will be measured, indicating $m=1$ was sent from the sender.
Otherwise, $m=0$ was transmitted from the sender.

For example, take $d=6$ and $N=8$,
the instruction access
sequences when sending $m=1$ and $m=0$ are as~follows:

\begin{itemize}
 \item \textbf{Init}: access instruction mix blocks $1 - 6$ mapping to set $x$
 \item \textbf{Encode}: access instruction mix blocks $7 - 9$ mapping to
 set $x$ (if $m=1$); $7 - 9$ of set $y$ (if $m=0$)
 (Stealthy) / no access (Fast) 
 \item \textbf{Decode}: access instruction mix blocks $1 - 6$
 mapping to set $x$ (if $m=1$, DSB and MITE are used; if $m=0$, LSD access is used)
\end{itemize}

\subsection{Non-MT Misalignment-Based Attack without Multi-Threading}
\label{subsec:single_misalign}

Similar to eviction-based non-MT attack shown in Section~\ref{subsec:singlethread}, misalignment 
can also be used to generate interference without multi-threading.
Details of this attack are not provided due to limited space.

For example, take $d=5, N=8, M=8$, the instruction access sequences
when sending $m=1$ and $m=0$ are as~follows:

\begin{itemize}
 \item \textbf{Init}: access instruction mix blocks $1 - 5$ mapping to set $x$
 \item \textbf{Encode}: access \textit{misaligned} instruction mix blocks $6 - 8$ mapping to set $x$ (if $m=1$); 
 \textit{aligned} instruction mix blocks $6 - 8$ mapping to set $x$
  (Stealthy) / no access (Fast) (if $m=0$)
 \item \textbf{Decode}: access instruction mix blocks $1 - 5$ mapping to set $x$ (if $m=1$, DSB 
 access is used; if $m=0$, LSD access is used)
\end{itemize}

\subsection{Slow-Switch Attack without Multi-Threading}
\label{subsec:slow_switch}

We now also present a covert-channel attack making use of LCP instructions, which we call the slow-switch attack.
For slow-switch attack, the receiver (attacker) starts and ends the timer
in the Init and Decode Steps.
Meanwhile, in the Encode Step,
within the loop, there will be in total $r$ number of LCP instructions being executed and the number of loops 
is $p$ (or $q$, $p=q$ as the same setting for non-MT eviction-based attacks).
When sending $m=1$, the sender will alternatively execute one normal \textit{add} instruction followed by 
one \textit{add} instruction with length changing prefix; this is repeated for $r$ times. This new type of instruction mix can enlarge the LCP stall cycles and maximize the LSD-to-DSB switches. 
When sending $m=0$, the sender will execute $r$ normal \textit{add} instructions and then execute $r$
\textit{add} instruction with length changing prefixes. This instruction mix has fewer LCP stalls,
thus minimizing the LSD-to-DSB switch penalties. 

For example, take $r=16$, the instruction access sequences when sending $m=1$ and $m=0$ are as~follows:

\begin{itemize}
 \item \textbf{Init}: start the timer.
 \item \textbf{Encode}: access $r=16$ groups of instructions, where each group has an \textit{add} instruction with a length changing prefix and then a normal \textit{add} instruction (if $m=1$); or
 access $16$ normal \textit{add} instruction and then $16$ \textit{add} instruction with length changing prefixes (if $m=0$); 
 \item \textbf{Decode}: stop the timer.
\end{itemize}

  \section{Evaluation of Timing-Channel Attacks}
\label{section_evaluation}

 \begin{figure}[!t]
 \centering
 \includegraphics[width=6.6cm]{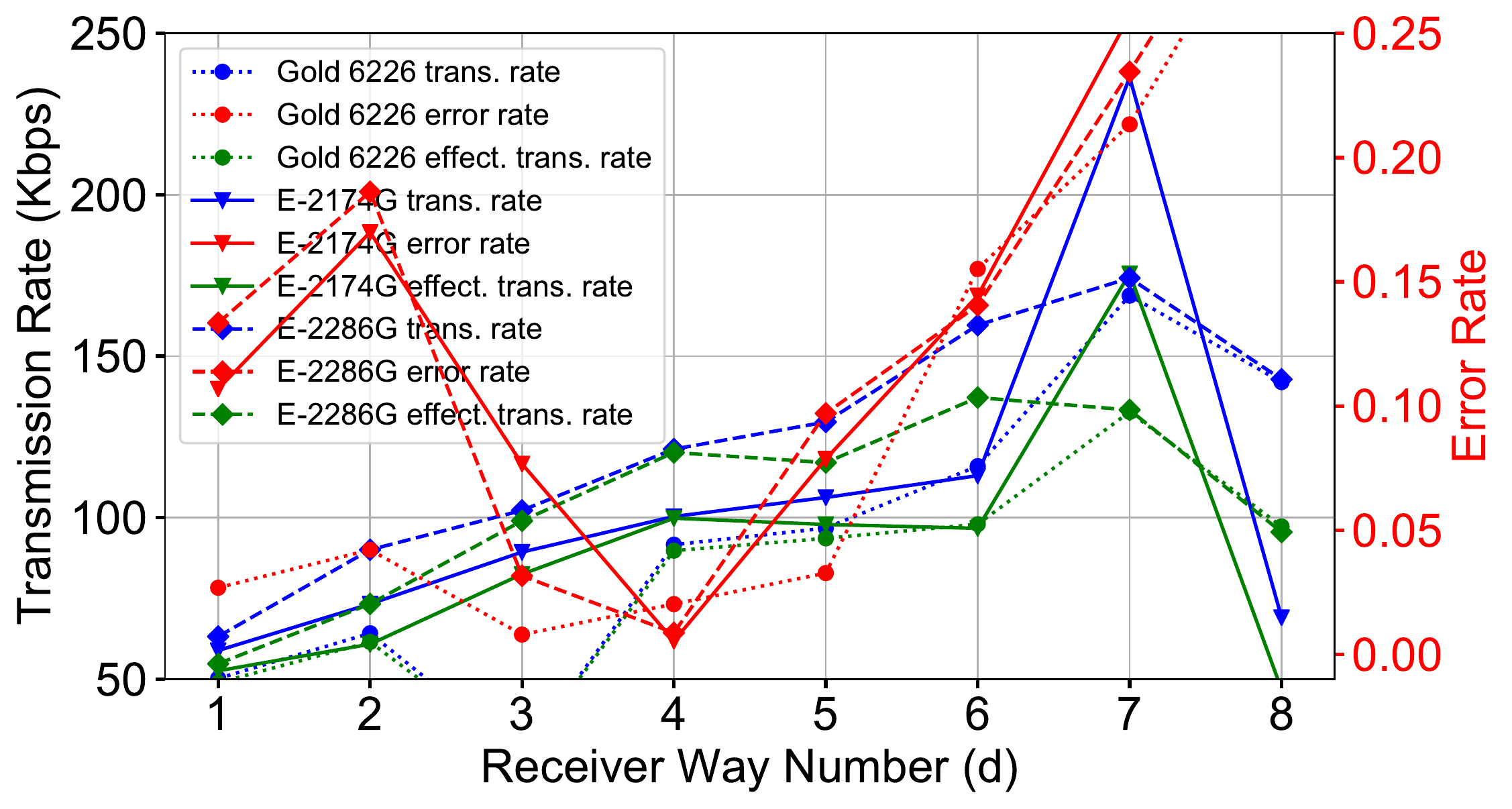} 
 \caption{\small 
 Evaluation of MT Eviction-Based Attack for different values of parameter $d$.
 }
 \label{fig:eviction}
 \end{figure}

 \begin{table*}[t]
 \caption{\small Transmission rates and error rates of the covert-channel MT Eviction-Based Attack when setting $d=1$ for 
 for different message patterns: all $0$s, all $1$s, alternating $0$s and $1$s, and random.
 }
 \label{tbl:eval_config}
 \centering
 \fontsize{6.7pt}{8.7pt}\selectfont
 \begin{tabular}{|p{0.62in}|p{0.31in}|p{0.346in}|p{0.346in}||p{0.31in}|p{0.346in}|p{0.346in}||p{0.31in}|p{0.346in}|p{0.346in}||p{0.31in}|p{0.346in}|p{0.346in}|}
 \hline
 
 & \multicolumn{3}{c||}{\textbf{{All $0$s Message}}} & \multicolumn{3}{c||}{\textbf{{All $1$s Message}}} 
 & \multicolumn{3}{c||}{\textbf{{Alternating $0$s and $1$s Message}}} & \multicolumn{3}{c|}{\textbf{{Random Message}}} \\ \cline{2-13}
 
 & G-6226 & E-2174G & E-2286G & G-6226 & E-2174G & E-2286G & G-6226 & E-2174G & E-2286G & G-6226 & E-2174G & E-2286G \\ \hline
 
 \textbf{{Tr. Rate (Kbps)}} & 42.66 & 49.53 & 87.33 & 55.28 & 61.17 & 102.39 & 50.21 & 58.86 & 64.96 & 18.28 & 21.80 & 25.61 \\ \hline
 \textbf{{Error Rate}} & 0.00\% & 0.00\% & 0.00\% & 0.00\% & 0.00\% & 0.00\% & 2.68\% & 10.69\% & 12.56\% & 22.57\% & 18.53\% & 19.83\% \\ \hline
 
 \end{tabular}
\end{table*}

\begin{table*}[t]
 \caption{\small 
 Transmission rates and error rates of all the eviction-based and misalignment-based attacks when setting 
 $d=6$ for eviction-based attacks and $d=5$, $M=8$ for misalignment-based attacks. 
 The transmitted message is alternating pattern of $0$s and $1$s.
 Transmission rates for the fastest attack are shown in bold.
 Intel Xeon E-2288G machine we tested has hyper-threading disabled so there is no MT attack possible.
 }
 \label{tbl:eval_covert}
 \centering
 \fontsize{6.7pt}{8.7pt}\selectfont
 \begin{tabular}{|p{0.62in}|p{0.23in}|p{0.23in}|p{0.23in}|p{0.23in}||p{0.23in}|p{0.23in}|p{0.23in}|p{0.23in}||p{0.23in}|p{0.23in}|p{0.23in}|p{0.23in}|}
 \hline
 & \multicolumn{4}{c||}{\textbf{{Non-MT Stealthy Eviction-Based}}} 
 & \multicolumn{4}{c||}{\textbf{{Non-MT Stealthy Misalignment-Based}}} & \multicolumn{4}{c|}{\textbf{{MT Eviction-Based}}} \\ \cline{2-13}

 & G6226 & 2174G & 2286G & 2288G & G6226 & 2174G & 2286G & 2288G & G6226 & 2174G & 2286G & 2288G \\ \hline
 
 \textbf{{Tr. Rate (Kbps)}} & 419.67 & 851.81 & 1182.55 & 1356.43 & 713.01 & 466.02 & 723.15 & 1094.39 & 115.97 & 113.02 & 161.63 & --- \\ \hline
 \textbf{{Error Rate}} & 6.48\% & 3.43\% & 3.45\% & 0.36\% & 22.56\% & 11.34\% & 16.56\% & 10.08\% & 15.52\% & 14.44\% & 13.93\% & --- \\ \hline\hline

 & \multicolumn{4}{c||}{\textbf{{Non-MT Fast Eviction-Based}}} & \multicolumn{4}{c||}{\textbf{{Non-MT Fast Misalignment-Based}}}
 & \multicolumn{4}{c|}{\textbf{{MT Misalignment-Based}}} \\ \cline{2-13}

 & G6226 & 2174G & 2286G & 2288G & G6226 & 2174G & 2286G & 2288G & G6226 & 2174G & 2286G & 2288G \\ \hline
 
 \textbf{{Tr. Rate (Kbps)}} & 501.06 & 977.68 & 1205.90 & 1399.96 & 500.90 & 959.45 & 1228.35 & \textbf{1410.84} & 129.36 & 152.44 & 200.37 & --- \\ \hline
 \textbf{{Error Rate}} & 6.09\% & 0.00\% & 0.00\% & 0.00\% & 0.16\% & 0.00\% & 0.16\% & \textbf{0.00\%} & 7.85\% & 2.77\% & 4.62\% & --- \\ \hline
 
 \end{tabular}
 \end{table*}
 
In this section, we evaluate the transmission rates and error rates of all the timing-based covert-channel attacks 
discussed in Section~\ref{sec:impl}. Power attacks, SGX attacks, use of new covert channels
in Spectre, microcode patch fingerprinting, and new side-channel attack are evaluated later.

The evaluation is conducted on $4$ recent $x86\_64$ processors from Intel
Skylake's family. The specifications of the processors is shown
in Table~\ref{tbl:CPU_info}.
For each covert channel, the transmitted data is compared with the received data
to compute the error rates.
To evaluate the error rates of the channel,
the Wagner-Fischer algorithm~\cite{navarro2001guided} is used to calculate the 
edit distance between the sent string and the received string.

\subsection{Number of Iterations ($p, q$) for Attack Steps}

After careful tuning of the configurations,
when sending each bit $m$ of message,
non-MT attacks can have $p=q=10$ (to repeat initialize, encode, and decode steps
and still reliably observe result with low error rates). To transmit each bit, the sender does one encoding step
and receiver does one decoding step and this pattern of activity is repeated in total $10$ times, hence $p=q=10$.
For MT attacks, for each bit to be transmitted the receiver does $10$ decoding measurements for each encoding step, while each
encoding step has to be repeated $100$ times, hence $p/q=10$, where $q=100$ (total encoding steps), $p=1000$ (total decoding steps).
The $q=100$ is due to more noise in the MT setting, compared to $q=10$ for the non-MT setting. 

\subsection{Threshold for Detecting Transmitted Bit}

To establish decoding threshold for timing measurements, to determine $m = 1$ vs. $m = 0$, an alternating pattern of $0$s and $1$s is sent, 
and the timing (measured in cycles using the {\tt rdtscp} instruction) is averaged for $0$s and $1$s to establish the threshold. 
Based on different covert channels, 
if a measurement is $30-70$\% or more above the threshold, it is judged to be a ``1'', otherwise it is judged to be a ``0''.
The simple encoding can be in future replaced with other channel coding methods~\cite{massey1978foundation}
for possibly faster transmission.

\subsection{Influence of ($d, M$) Parameters}

To help find the ideal transmission rate, we evaluate the influence of $d$ (number of DSB ways accessed by the receiver) 
and its impact on the transmission rate and error rates.\footnote{This 
work is not aimed at achieving the highest bandwidth covert channel.
To fully optimize the transmission rate and error rate, techniques such as the ones used 
in~\cite{saileshwar2021streamline} can be further 
exploited. }
The results of changing $d$ for
MT Eviction-Based Attack is
shown in Figure~\ref{fig:eviction}. 
When increasing $d$ from $1$ to $8$ (DSB has $N=8$ ways), 
the number of ways accessed by the sender will decrease (number of sender's ways accessed is $N+1-d$).
Receiver's observation will then become less stable (error rate increases) while 
on the other hand 
transmission rate increases.
Error rates of small $d$ (e.g., $d=1,2$) are also large because 
when the number of ways accessed by the receiver is small, timing difference of sending 
$0$ and $1$ is small, which can be disrupted by the system noise.
To find a balance between the transmission rate and error rate, we choose $d=6$ for 
eviction-based attacks.
For misalignment-based attacks, we choose 
$d=5$, $M=8$ ($M$ is the total number of ways accessed by the sender and receiver for misalignment-based attacks).

\subsection{Influence of Message Patterns}

A sample evaluation of MT Eviction-Based Attack for the four different message patterns with $d=1$
is shown in Table~\ref{tbl:eval_config}.
From the results it can be seen that better transmission rate and error rate are derived for all $0$s and all $1$s.
This is possibly because when not changing the bits (as is case for all $0$s or all $1$s),
the frontend path used by the sender accesses remains the same, generating less noise.
The random messages are the worst due to the frequent and unstable frontend path changes.

 \begin{table}[t]
 \caption{\small Transmission rates and error rates of Slow-Switch Attacks.
 The transmitted message is alternating $0$s and $1$s.}
 \label{tbl:lcp_switch}
 \centering
 \fontsize{6.7pt}{8.7pt}\selectfont
 \begin{tabular}{|c|C{0.45in}|C{0.45in}|}
 \hline
 & \multicolumn{2}{c|}{\textbf{{Non-MT Slow-Switch-Based}}} \\ \cline{2-3}
 & G6226 & 2288G \\ \hline
 \textbf{{Tr. Rate (Kbps)}} & 678.11 & 1351.43 \\ \hline
 \textbf{{Error Rate}} & 6.74\% & 0.64\% \\ \hline
 \end{tabular}
 \end{table}

\subsection{Transmission Rates and Error Rates}

The bit transmission rates and error rates for all types of the timing-based attacks are presented in Table~\ref{tbl:eval_covert} 
and Table~\ref{tbl:lcp_switch},
with
$d=6$ for eviction-based attacks, $d=5$ for misalignment-based attacks and $r=16$ for slow-switch attacks.
For the best attack, which is the Non-MT Fast Misalignment-Based Attack, the
transmission rate can be as high as $1410$ Kbps ($1.41$ Mbps) with almost 0\% error rate.
Slow-switch attacks have generally similar transmission rate compared with the non-MT misalignment-based attacks.
Non-MT attacks have better transmission rate than MT attacks due to 
smaller noise.

  \section{Power-Channel Attack Evaluation}
\label{sec:power_attack}

Switching between LSD or DSB and the MITE will not only cause 
timing changes for instruction processing, but also power changes.
The power changes can be measured
by abusing unprivileged access to Intel's Running 
Average Power Limit (RAPL) interface~\cite{gough2015energy}.%
\footnote{
In power attacks, if unprivileged RAPL accesses are prevented,
we can still use privilege access and use power 
to attack SGX enclaves. We do not show this type of attack due to the limited space.
}

\begin{figure}[!t]
 \centering
 \includegraphics[width=4cm]{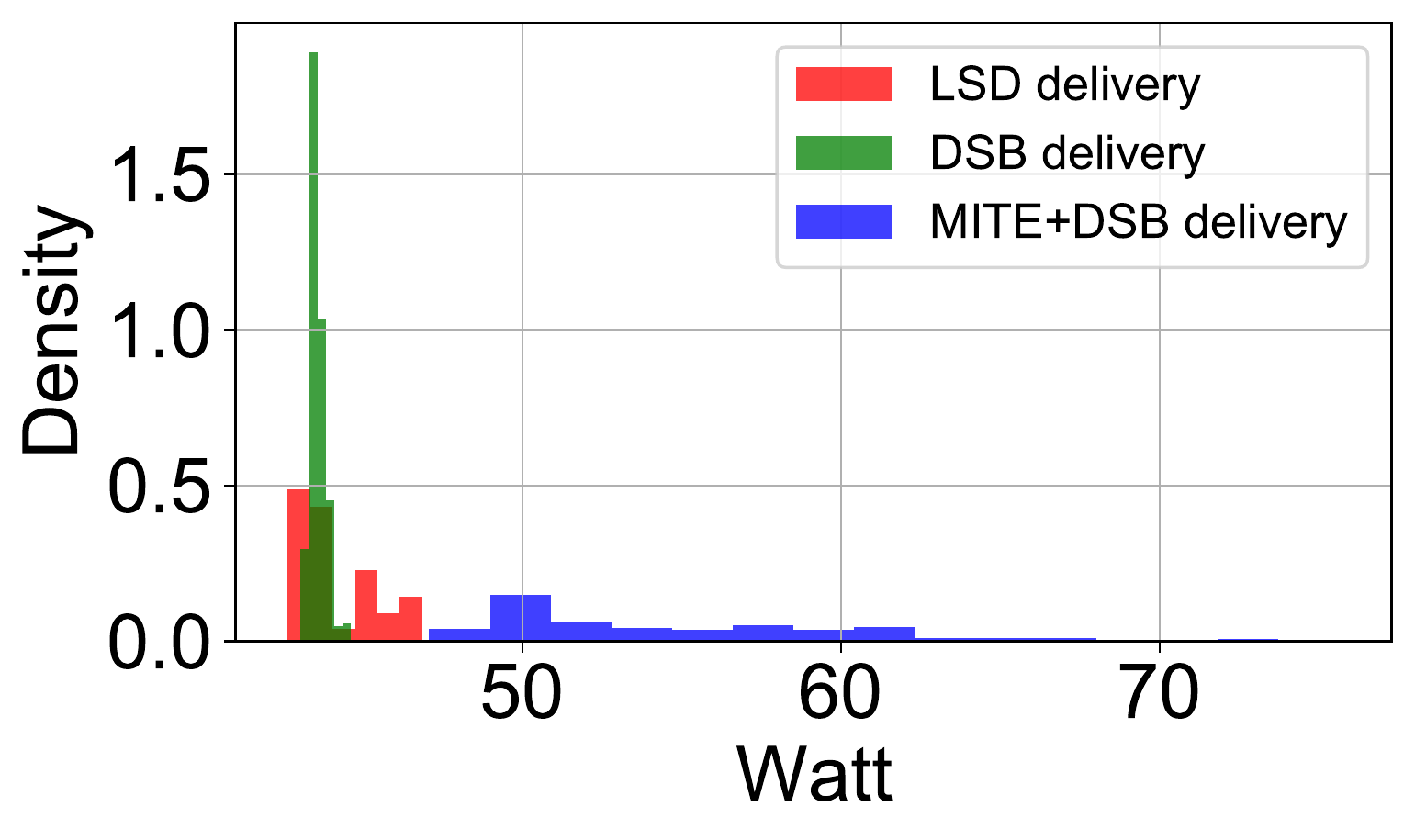}
 \caption{\small Example histogram of power consumption when different frontend paths are
 used to process micro-ops in Intel Xeon Gold 6226~processor.}
 \label{fig:power_1}
\end{figure}

Figure~\ref{fig:power_1} shows example histogram of the power consumption
of utilizing different frontend paths for the micro-ops 
in Intel Xeon Gold 6226~processor. Based on the power differences, we demonstrate
a non-MT attack that can detect LSD or DSB vs. MITE frontend path power differences
caused by eviction or misalignment
through
observing the power changes in RAPL.
Configuration of the attack is similar to the non-MT attack demonstrated in
Section~\ref{subsec:singlethread}.
To observe the power differences, for each bit transmission the initialize, 
encode, and decode steps have to be iterated for
$p=q=240,000$ times since RAPL interface update interval is 
around $20$kHz~\cite{lipp2021platypus}. The power attack's bandwidth is limited
by the update interval of RAPL, and is less than for the timing attacks.

Table~\ref{tbl:power} shows the evaluation results of
two power-based non-MT attacks on Intel's Xeon Gold 6226~processor.
The bandwidth of the power attacks
is around $0.6$ -- $0.7$ Kbps.
The transmission is still above $100$ bps which is considered a high-bandwidth channel
by TCSEC~\cite{orangebook}.
The power attack bandwidth can possibly be further improved 
using techniques such as the ones shown in recent PLATUPUS work~\cite{lipp2021platypus}.

\begin{table}[t]
 \caption{\small Evaluation of Non-MT Power-Based attacks on 
 Intel Xeon Gold 6226~processor when setting $d=6$.
 }
 \label{tbl:power}
 \centering
 \scriptsize 
 \begin{tabular}{|c|c|c|} 
 \hline
 & \textbf{{Eviction-Based}} & \textbf{{Misalignment-Based}} \\ \hline

 \textbf{{Tr. Rate (Kbps)}} & 0.66 & 0.63 \\ \hline
 \textbf{{Error Rate}} & 18.87\% & 9.07\% \\ \hline 

 \end{tabular}
\end{table}

  \section{SGX Attack Evaluation}
\label{sec:sgx}

 \begin{table*}[t]
 \caption{\small Transmission rates and error rates of 
 covert channels for leaking information from an SGX enclave when setting $d=6$ 
 for eviction-based attacks and $d=5$, $M=8$ for misalignment-based attacks. 
 The transmitted message is alternating $0$s and $1$s.
 Intel Xeon E-2288G machine we tested has hyper-threading 
 disabled so no MT attack data is provided for this machine.
 }
 \label{tbl:eval_sgx1}
 \centering
 \fontsize{6.7pt}{8.7pt}\selectfont
 \begin{tabular}{|p{0.62in}|p{0.34in}|p{0.34in}|p{0.34in}||p{0.34in}|p{0.34in}|p{0.34in}||p{0.34in}|p{0.34in}|p{0.34in}|}
 \hline

 \multirow{2}{*}{SGX Attacks} & \multicolumn{3}{c||}{\textbf{{Non-MT Stealthy Eviction-Based}}}
 & \multicolumn{3}{c||}{\textbf{{Non-MT Stealthy Misalignment-Based}}} & \multicolumn{3}{c|}{\textbf{{MT Eviction-Based}}} \\ \cline{2-10}
 
 & E-2174G & E-2286G & E-2288G & E-2174G & E-2286G & E-2288G & E-2174G & E-2286G & E-2288G \\ \hline
 
 \textbf{{Tr. Rate (Kbps)}} & 18.96 & 19.56 & 21.20 & 23.93 & 24.70 & 27.10 & 7.85 & 14.89 & --- \\ \hline
 \textbf{{Error Rate}} & 0.16\% & 1.33\% & 2.18\% & 0.32\% & 0.76\% & 0.76\% & 6.74\% & 8.02\% & --- \\ \hline
 
 \multirow{2}{*}{SGX Attacks} & \multicolumn{3}{c||}{\textbf{{Non-MT Fast Eviction-Based}}} 
 & \multicolumn{3}{c||}{\textbf{{Non-MT Fast Misalignment-Based}}} & \multicolumn{3}{c|}{\textbf{{MT Misalignment-Based}}} \\ \cline{2-10}
 
 & E-2174G & E-2286G & E-2288G & E-2174G & E-2286G & E-2288G & E-2174G & E-2286G & E-2288G \\ \hline
 
 \textbf{{Tr. Rate (Kbps)}} & 29.35 & 32.01 & 34.48 & 30.36 & 31.18 & 35.20 & 6.39 & 13.62 & --- \\ \hline
 \textbf{{Error Rate}} & 0.04\% & 1.40\% & 0.40\% & 0.08\% & 1.08\% & 0.68\% & 2.56\% & 12.95\% & ---\\ \hline

 \end{tabular}
 \end{table*}

The goal of Intel Software Guard Extension (SGX) is to protect sensitive data 
against the untrusted user, even on already compromised system, 
with the help of hardware-implemented security and cryptographic mechanism 
inside the processor~\cite{IntelManual}.
Unfortunately, as we demonstrate, SGX is also vulnerable to frontend-related attacks.%
\footnote{We demonstrate attacks on SGX, although there is a 
newer SGX2 which extends SGX with dynamic memory management and other features,
we believe these features will not affect our attacks and our attacks 
can be applied to SGX2 in future when machines with SGX2 are~available.}

To demonstrate our attacks
in an SGX environment, we assume a sender program is
running inside the SGX enclave and manipulates the use of the frontend paths
to communicate to a receiver outside of the~SGX. 
We consider both non-MT and MT SGX attacks, but for both there is 
only one SGX entry and one SGX exit, while attacker measures
the execution time from the outside.
Consequently, instruction TLB flushing upon entry and exit does not impact our attacks. 

\subsubsection{MT Timing-Based SGX Attacks}

For MT timing-based SGX attacks,
the sender maintains its own thread and performs the covert transmission from within the enclave.
Meanwhile, the receiver decodes bits of the sender by measuring the timing of its own operations. 
Under this scenario, 
the receiver is able to detect the performance difference 
of its own instruction access based on the activity inside the SGX.
If SGX thread is running, then the receiver will observe the partitioned DSB. 
If the SGX thread is idle, whole DSB is dedicated
to the receiver thread. Receiver can observe its own internal-interference and deduce the DSB~state.

Evaluation of the MT timing-based SGX attacks is shown in Table~\ref{tbl:eval_sgx1}.
It can be seen from the table that the transmission rates of SGX attacks 
can be roughly 
$6$ Kbps -- $15$ Kbps
with iteration numbers 
$p=1,000$, $q=10,000$,
while maintaining the similar error rates as the MT non-SGX attacks.

\subsubsection{Non-MT Timing-Based SGX Attacks}

For non-MT timing-based SGX attacks,
the sender program is still inside the
enclave, while 
the receiver derives the information by measuring the timing of SGX operation 
from outside of the enclave.
Under this scenario, the receiver's observations depend on ability to detect the internal
interference of the sender's accesses within the enclave,
to detect whether
there are frontend path changes caused by the eviction or misalignment of the micro-ops or not.
The non-MT SGX attacks, because they do not leverage multi-threading, are possible even when multi-threading is disabled for security.

In the non-MT setup, we assume the attacker (receiver) is able to trigger the sender and they both
execute on the same hardware thread.
To reduce overhead and noise of enclave exits and entrances, for each transmission of a bit,
there is only one entrance and exit. Effectively the receiver starts time 
measurement, then allows the enclave to run,
and then finally measures the timing of the enclave as it was affected by the frontend paths.
Compared to non-SGX attacks, more iterations of initialization, encoding, and decoding are necessary 
($p=q=1,000-5,000$ iterations for the SGX attack compared
to $p=q=10$ iterations for non-SGX attacks)
in order to transmit one bit.

Evaluation of the non-MT timing-based SGX attacks is shown in Table~\ref{tbl:eval_sgx1}.
As the table shows, the transmission rates of non-MT SGX attacks are 
roughly $1/25$ to $1/30$ of non-MT non-SGX 
attacks, while still maintaining acceptable and even lower error rates.

\subsubsection{Power-Based SGX Attacks}

Power-based attacks are also possible, but not discussed due to limited space. We remark, however,
that even if RAPL is disabled for user-level code, power-based SGX attacks are possible because RAPL 
can be accessed from the privileged, malicious OS.

  \section{Frontend and Instruction Cache-Based Spectre Attack Evaluation}
\label{sec:spectre_attack}

 \begin{table}[t]
 \caption{\small L1 miss rates of our Spectre v1 version attack (run on Intel's Xeon Gold 6226 processor) 
 with variants of Spectre v1 that use different covert channels. 
 MEM F+R, L1D F+R, and L1D LRU attacks are from work~\cite{xiong2020leaking}. L1 miss rates 
 in~\cite{xiong2020leaking} are L1D miss rates.}
 \label{tbl:eval_spectre}
 \centering
 \fontsize{6.7pt}{8.7pt}\selectfont
 \begin{tabular}{|p{0.52in}|p{0.28in}|p{0.28in}|p{0.28in}|p{0.17in}|p{0.17in}|p{0.17in}|}
 \hline
 & \multicolumn{3}{c|}{\textbf{{Others}}} & \multicolumn{3}{c|}{\textbf{{Our}}} \\
 \hline
 & \textbf{{MEM F+R}}~\cite{xiong2020leaking} & \textbf{{L1D F+R}}~\cite{xiong2020leaking} 
 & \textbf{{L1D LRU}}~\cite{xiong2020leaking} & \textbf{{L1I F+R}} 
 & \textbf{{L1I P+P}} & \multicolumn{1}{c|}{\textbf{{Frontend}}} \\ \hline

 \textbf{{L1 Miss Rate}} & 2.81\% & 4.79\% & 4.48\% & 0.45\% & 0.48\% & 0.21\% \\ \hline
 
 \end{tabular}
 \end{table}

Speculative attacks leverage transient
execution to access secret and then a covert channel to pass
the secret to the attacker~\cite{Kocher2018spectre,lipp2018meltdown,canella2019systematic}. 
In this section, we demonstrate our new variants of Spectre v1.
In our Spectre attacks, we assume an in-domain attack where the victim and attacker
code are in the same thread, so only one thread is running on the processor core.
The secret message is represented by $5$ bit chunks
(each chunk can have value from $0$ to $31$).
We then use one of the $32$ DSB sets to represent each value.
Similar to cache-based channels, during the speculative execution, secret value is encoded
by accessing the corresponding set. Unlike other cache attacks, to access a DSB set,
instruction mix block mapping to that set has to be executed.
We also implemented Spectre v1 attacks using L1I cache Flush + Reload attack and L1I
Prime + Probe attack, to compare to our frontend attacks.

Table~\ref{tbl:eval_spectre} shows the L1 miss rate when using our channels compared to other channels.
While our
Spectre v1 attacks have lower bandwidths than data cache-based Spectre attacks,
we are able to achieve lowest L1 miss rates.
Especially, compared with recent cache-based LRU~\cite{xiong2020leaking} covert channels which target stealthy 
attacks without causing high data cache miss rates,
our frontend attack does not cause any cache misses at all,
making the L1 miss rate the~smallest.

  \section{Microcode Patch Detection Evaluation}
\label{sec:patch}

\begin{figure}
 \centering
 \includegraphics[width=\linewidth,height=3.2cm]{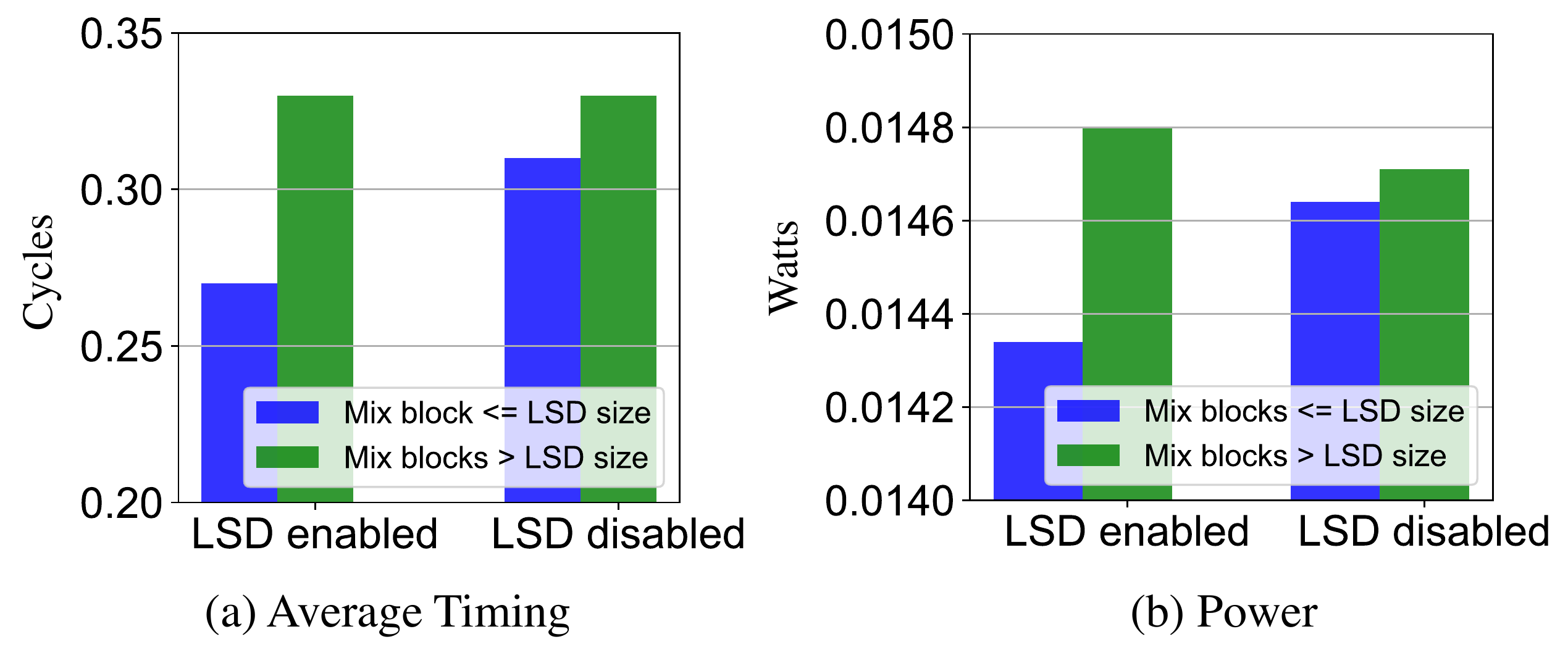}
 \caption{Example comparison of frontend timing and power for executing
 instruction mix blocks less or greater than LSD capacity. All mix blocks map to 
 the same DSB set. If LSD is disabled execution falls back to DSB and MITE.}
 \label{fig:patch}
 \end{figure}

When evaluating the behavior of the processor frontend, we also found a new type of attack
where performance of the frontend can be used for 
fingerprinting the microcode updates of the processor.
In particular, we evaluated our Intel Xeon Gold 6226 test machine under older {\tt 3.20180312.0ubuntu18.04.1} 
(patch1) and newer {\tt 3.20210608.0ubuntu0.18.04.1} (patch2)
Intel microcode patches. While neither patch explicitly mentions LSD, we found that with the newer
patch2
LSD is disabled while with older
patch1 the LSD is enabled. To switch between the patches,
the processor has to be restarted so the microcode in the CPU can be updated.

To detect the changes in the LSD behavior, we can use both the timing difference and the power 
difference when testing code sequences with number of instruction mix blocks less than LSD capacity 
(so they would fit in LSD and be processed by LSD)
or sequences with number of instruction mix blocks greater than LSD capacity 
(so micro-ops would be forced to be handled by DSB and MITE instead).
The average timing and power difference for LSD enabled (patch1) vs.
disabled (patch2)
are shown in Figure~\ref{fig:patch}. Attackers can clearly differentiate which patch has been applied,
with timing being a more reliable indicator.

Attackers can leverage this to learn of vulnerabilities of the processor. For example,
patch2 
protects against CVE-2021-24489: potential security vulnerability in some Intel Virtualization Technology for Directed I/0 (VT-d) products
that allows for escalation of privilege.%
\footnote{
The patch2 also
adds protections
against CVE-2021-24489, CVE-2020-24511, CVE-2020-24512, and CVE-2020-24513.
}
Knowing the patch is applied or not allows the attacker to exploit VT-d related attacks.
The frontend timing thus cannot only be the target of attack itself, 
but help attacker discover other vulnerabilities in the system.

  \section{Evaluation of Side-Channel Attack and Fingerprinting of Applications}
\label{sec_fingerprinting}

\begin{figure*}
 \centering
 \begin{minipage}{0.75\textwidth}
 \centering
 \includegraphics[height=3cm]{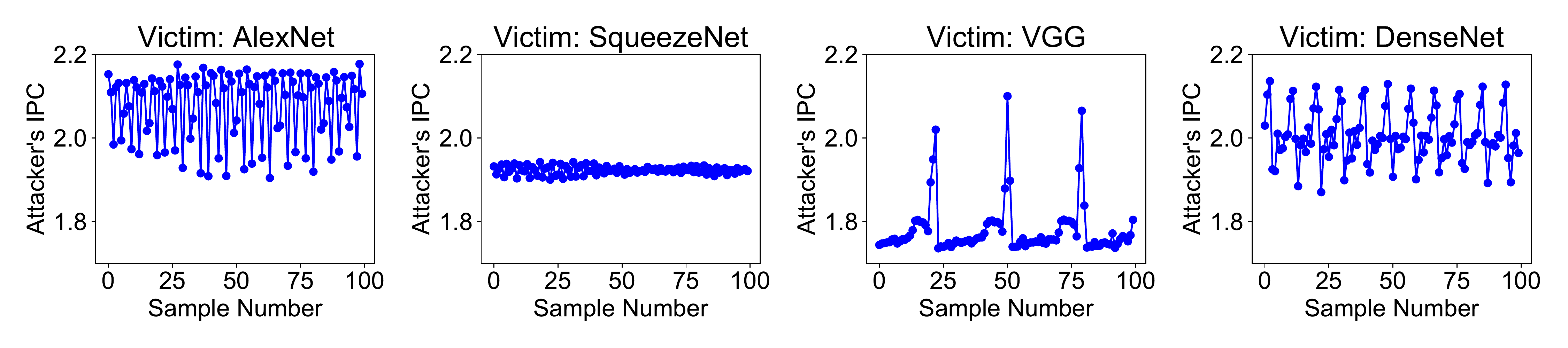}
 \caption{\small Fingerprinting results of machine learning model using frontend side-channel attacks.
 Baseline IPC of the attacker program is $3.58$. With two threads the IPC is roughly halved. Furthermore, due to
 different patterns of the victim it fluctuates between the $1.8$ and $2.2$.
 }
 \label{fig:cnn}
 \end{minipage}\hfill
 \begin{minipage}{0.22\textwidth}
 \centering
 \includegraphics[height=3.2cm,width=4cm]{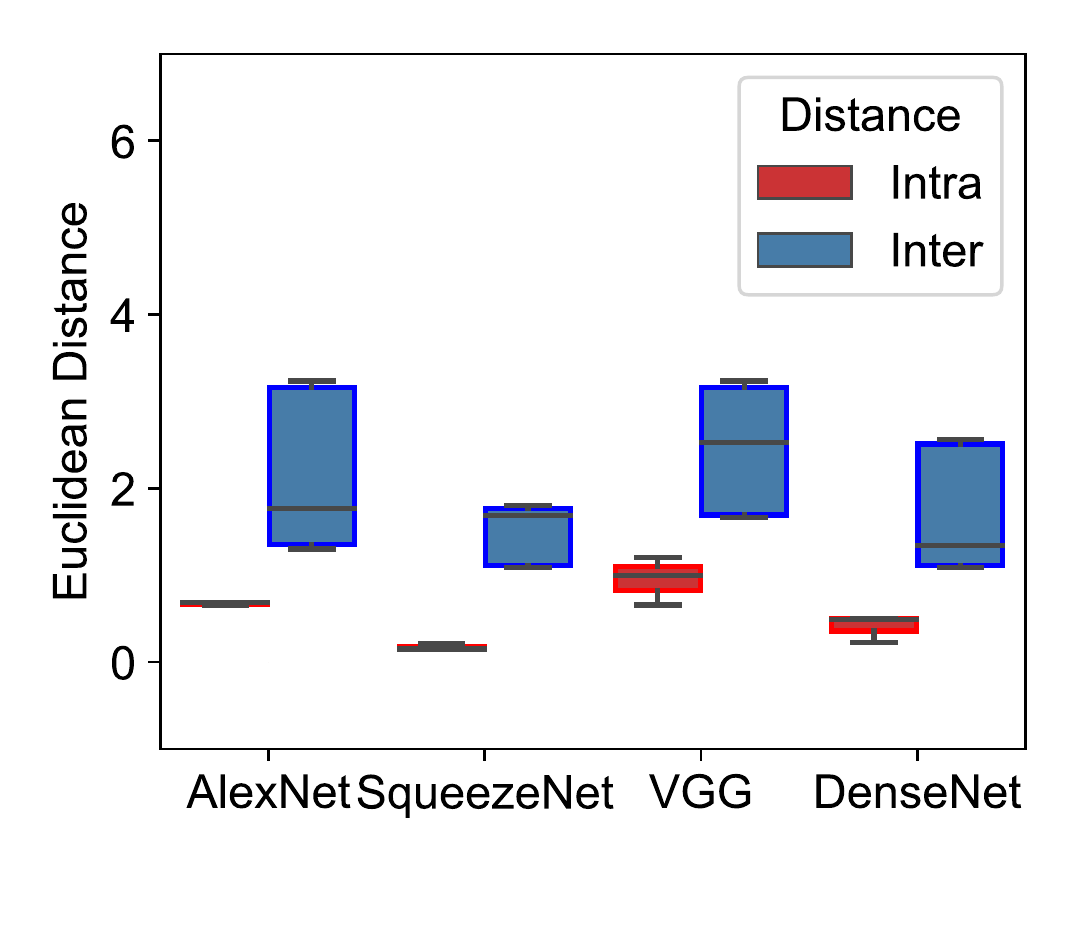}
 \caption{\small Inter-distance and intra-distance of all the models.}
 \label{fig:cnn_d}
 \end{minipage}
\end{figure*}

Based on the frontend characteristics, we developed a 
new frontend-based fingerprinting technique utilizing a side-channel attack 
to demonstrate that frontend can be not only used for covert communication, 
but also for side-channel information leakage.
Our fingerprinting technique is able to identify what type of workload 
a victim is running on a co-located SMT thread.
Moreover, our technique can achieve fingerprinting using low-frequency timing measurements, 
therefore, it works on platforms where access to high-precision timers is limited.
The approach does not use any performance counters or privileged access, 
and depends only on the attacker (receiver) measuring their own instructions per cycle (IPC). 
The IPC is affected by the shared frontend, especially the shared MITE, 
and interference between attacker and victim in the frontend are the sources of the information leakage. 
The attacks were tested on same CPUs as the covert channels and work with current Intel processors 
where DSB and LSD are partitioned between threads (but MITE is not).

When compared with previous fingerprinting techniques~\cite{shusterman2019robust,oren2015spy}, 
which are mostly based on using performance counters or contention in the backend of the processor,
our side-channel attack has number of advantages.
Our method 1) does not need to measure the performance of the victim workload, 
2) does not require usage of any performance counters but only a low-precision timer,
3) does not depend on eviction of lines in instruction and data caches
so it is robust against the existing defense measures on caches, and 4) 
it is also robust against existing frontend resource 
hardware partitioning, including DSB partitioning and LSD partitioning implemented on Intel microarchitectures. 

\subsection{Side Channel Design}

To develop the side channel, we designed a modified receiver that 
uses a new mix block of \textit{nop} instructions instead of the prior instruction mix blocks used in the covert channels.
We use \textit{nop} instructions 
in the x86 ISA to construct our attacker thread, which naturally triggers frontend resources to decode the \textit{nops}, 
but it does not generate any traffic in the backend. The attacker thread used to perform fingerprinting 
loops through $100$ \textit{nop} instructions which will not fit in LSD but are able to fit in DSB.
The loop takes two cache lines,
which never get evicted from the cache because of the repeated loop access
within the attacker program.
Victim program will slow down the decoding process of the MITE for the attacker which causes timing variation
of the attacker program, and when the attacker measures its own performance 
variation, it is able to observe patterns that reveal type of victim application. 
The attacker measures its own performance by computing the IPC based on the number of \textit{nops} 
executed and time reading from the \textit{rdtsc}.

We measure only the instruction per second at a low frequency of 10Hz because existing
platforms limit the usage of high-precision timers~\cite{shusterman2019robust}. 
Euclidean distance~\cite{danielsson1980euclidean} is used to calculate the distance of 
IPC measurement traces of two test results.
If these two tests of the attacker program run with the same victim benchmark, 
intra-distance is derived. Otherwise, inter-distance is derived.
Furthermore, we verified that the contention indeed happens in the frontend by monitoring the 
performance counter changes. 
Note that the actual attack does not use performance counters. They were only used to validate the results.

\subsection{Fingerprinting of Mobile Applications}

To demonstrate the fingerprinting and the side-channel attack on mobile application usage, 
we performed the experiments 
using a popular
Geekbench5 benchmark suite~\cite{geekbench5}.
It consists of a 
wide range of workloads including camera, navigation, 
speech recognition, etc.

We run the attacker thread along with a Geekbench5 thread on a single SMT-enabled core.
Unique IPC waveforms of the attacker are derived when running with different benchmarks.
We observe an average $0.232$ intra-distance vs. $4.793$ inter-distance for the $10$ benchmarks tested. Our results indicate that the
IPC changes of the attacker thread can be used directly to distinguish the type of the victim~application that is running.

\subsection{Fingerprinting of Machine Learning Algorithms}

We also demonstrate the fingerprinting of different machine learning algorithms from the TVM framework.
Figure~\ref{fig:cnn} shows the average IPC traces
of the attacker program thread when running with different 
CNN model inference threads on the same SMT core. 
Clear differences in the
traces are shown and these can be used to distinguish different machine learning models based on
the traces using different convolution layers. A set of traces
can thus be compared to reference traces to distinguish a
network. 
Because of the frontend contention in the MITE, even with partitioned LSD and DSB, the attacker can leak 
information about type of victim machine learning model.
As can be seen in Figure~\ref{fig:cnn_d},
the inter distance and intra distance
can be clearly differentiated.
This shows that the fingerprinting results 
can clearly 
differentiate machine learning model architectures. 
We observe an average $0.550$ intra-distance vs. $1.937$ inter-distance
for tested $4$ CNN models.

  \section{Discussion}

The frontend vulnerabilities do not involve interference in traditional instruction or data caches,
and they do not involve speculation. Therefore, a large set of existing defense mechanism will not be able to 
prevent them~\cite{yan2018invisispec,qureshi2018ceaser,kiriansky2018dawg}.
The major difficulty of dealing with the security vulnerabilities of the frontend paths is that
the frontend is designed to give better performance or lower power for different execution scenarios,
which inevitably creates inherent timing or power signatures. 
Eliminating these timing or power signatures would reduce the performance or power benefits.
Since frontend components such as the MITE, DSB, and LSD are widely used in modern architecture designs. 
Defending the frontend vulnerabilities will require new approaches for the design of the frontend.

At the system-level, the SMT can be always disabled for security-critical applications, 
which would eliminate the MT attacks.
This should be probably already done due to other prior attacks on caches, for example.

Even with SMT disabled, the non-MT attacks are possible. Defending these
would require careful design of the code so that there is no secret-dependent timing.  This requires
writing of the code to make sure that the frontend switching or timing is always the same, regardless
of the secret data being processed.  Instruction alignment, as shown by our misalignment-based attacks, can also
cause timing differences, so not just the code, but its location in the address space needs to be considered.

Regarding Spectre attacks, the frontend state should not be updated due to speculative execution.
Existing defenses such as buffering cache updates could be applied to the DSB.

For power-based attack, the ability to monitor power of other users or SGX enclaves needs to be disabled.
For user-level code, existing patches from Intel should be applied to disable access to the power monitors.
For SGX, the power monitors can be enabled in debug mode for development, but disabled in production mode.

Since patch detection is based on timing observation of whether some components are enabled or disabled,
there does not seem to exist an easy solution (unless all frontend paths have same timing,
which defeats the purpose of having different paths to get better performance).
System administrator should assume that potential
attackers know exactly which patches have been applied, and the patch level of the system should not
be considered a secret.

Although a number of attacks have been demonstrated in our work, we do note that
to perform some of the attacks we need to find specific instruction mix blocks to 
minimize the contention in the backend to allow the attacks to be effective.
The attacks may be difficult to deploy in practice, for example, if the right instruction mix block is not available in the code.
Nevertheless, our other attacks such as the side-channel and application fingerprinting 
do not depend on specific instruction mix blocks, but overall operation of the victim program.  The frontend then
can impact the system security, and more evaluation of the defenses and how to deploy them are needed.

  \section{Conclusion}
\label{sec:conclusion}

This paper evaluated security vulnerabilities in the processor frontend.
The work demonstrated numerous threats due to the timing and power signatures of MITE, DSB, and LSD, including high bandwidth covert channels.
Further this paper showed an SGX attack, 
a version of Spectre attack, a new microcode fingerprinting approach, and a new side-channel attacks for fingerprinting
applications.  The evaluated security threats demonstrated that the processor frontend can be a security weakness in a system,
even if existing defenses for the backend components are deployed. As a result, processor designers should pay more attention
to the frontend and its impact on security.

\section*{Acknowledgement}
This work was supported in part by NSF grant \nsf{1813797}.
Shuwen Deng was supported through the Google PhD Fellowship.
We thank Abhishek Bhattacharjee for
his insightful comments and feedback.

\bibliographystyle{IEEEtranS}
\bibliography{refs}

\clearpage

\end{document}